\documentclass[floatfix,
aps,pra,
reprint,
superscriptaddress,
amsmath,
amssymb
]{revtex4-1}

\usepackage[utf8]{inputenc}
\usepackage{times}
\usepackage{microtype}
\usepackage{graphicx}
\usepackage{upgreek}
\usepackage{bm}
\usepackage{hyperref}
\usepackage{xcolor}
\usepackage[normalem]{ulem}
\usepackage{placeins}
\usepackage{amsmath}
\usepackage{amsthm}
\usepackage{dcolumn}
\usepackage{comment}
\usepackage{soul}
\usepackage{bbm}
\usepackage{braket}
\usepackage{simpler-wick}

\usepackage[T1]{fontenc}
\usepackage{csquotes}
\usepackage[german,ngerman,english]{babel}
\usepackage{centernot}
\usepackage{mathtools}
\usepackage{ifthen}
\usepackage{tikz}
\usetikzlibrary{calc}
\usetikzlibrary{decorations.pathreplacing}
\usetikzlibrary{decorations}
\usetikzlibrary{positioning}
\usepackage{xxcolor}
\definecolor{structure}{rgb}{0.23,0.4,0.7}
\usepackage[normalem]{ulem}

\newcommand{\nn}{\nonumber}
\newcommand{\ud}{{\textrm{d}}}
\newcommand\underrel[2]{\mathrel{\mathop{#2}\limits_{#1}}}

\newsavebox{\blocksavebox}

\definecolor{niceblue}{rgb}{0.33,0.5,0.8}

\newcommand{\refsub}[2]{\hyperref[#1]{\ref*{#1}#2}}

\newcommand{\norm}[2][]{
  \ifthenelse{\equal{#1}{}}
    {\left\| {#2} \right\|}
    {\ifthenelse{\equal{#1}{uinv}}
      {\left\vert\kern-0.25ex\left\vert\kern-0.25ex\left\vert {#2} \right\vert\kern-0.25ex\right\vert\kern-0.25ex\right\vert}
      {\left\| {#2} \right\|_{#1}}
    }
}

\newcommand{\taverage}[2][]{
  \ifthenelse{\equal{#1}{}}
  {\overline{#2}}
  {\overline{#2}^{#1}}
}

\newcommand{\tracedistance}[3][]{
  \ifthenelse{\equal{#2}{}}
  {\ifthenelse{\equal{#3}{}}
    {\mathcal{D}_{#1}}{}
  }{
    \ifthenelse{\equal{#1}{}}
    {\mathchoice{\operatorname{\mathcal{D}}\left(#2,#3\right)}{\operatorname{\mathcal{D}}(#2,#3)}{\operatorname{\mathcal{D}}(#2,#3)}{\operatorname{\mathcal{D}}(#2,#3)}}
    {\mathchoice{\operatorname{\mathcal{D}}_{#1}\left(#2,#3\right)}{\operatorname{\mathcal{D}}_{#1}(#2,#3)}{\operatorname{\mathcal{D}}_{#1}(#2,#3)}{\operatorname{\mathcal{D}}_{#1}(#2,#3)}}
  }
}
\newcommand{\fidelity}[3][]{
  \ifthenelse{\equal{#2}{}}
  {\ifthenelse{\equal{#3}{}}
    {\mathcal{F}_{#1}}{}
  }{
    \ifthenelse{\equal{#1}{}}
    {\mathchoice{\operatorname{\mathcal{F}}\left(#2,#3\right)}{\operatorname{\mathcal{F}}(#2,#3)}{\operatorname{\mathcal{F}}(#2,#3)}{\operatorname{\mathcal{F}}(#2,#3)}}
    {\mathchoice{\operatorname{\mathcal{F}}_{#1}\left(#2,#3\right)}{\operatorname{\mathcal{F}}_{#1}(#2,#3)}{\operatorname{\mathcal{F}}_{#1}(#2,#3)}{\operatorname{\mathcal{F}}_{#1}(#2,#3)}}
  }
}

\newcommand{\Sr}[3][]{
  \ifthenelse{\equal{#1}{}}
    {\operatorname{\mathnormal{S}}(#2\|#3)}
    {\operatorname{\mathnormal{S}}_{#1}(#2\|#3)}
}



\definecolor{jens}{rgb}{0.1,0.5,0.1}
\newcommand{\je}[1]{{\color{black} #1}}

\begin{document}

\title{Unraveling long-time quantum dynamics using flow equations}

\author{S.~J.~Thomson}
\address{Dahlem Center for Complex Quantum Systems, Freie Universit{\"a}t Berlin, 14195 Berlin, Germany}

\author{J.\ Eisert}
\address{Dahlem Center for Complex Quantum Systems, Freie Universit{\"a}t Berlin, 14195 Berlin, Germany}
\address{Helmholtz-Zentrum Berlin f{\"u}r Materialien und Energie, 14109 Berlin, Germany}

\begin{abstract}
The study of many-body quantum dynamics in strongly-correlated systems is extremely challenging. To date few numerical methods exist which are capable of simulating the non-equilibrium dynamics of two-dimensional quantum systems, in part reflecting complexity theoretic obstructions. In this work, we present a new technique able to overcome this obstacle, by combining continuous unitary flow techniques with the newly developed method of scrambling transforms. We overcome the prejudice that approximately diagonalizing the Hamiltonian cannot lead to reliable predictions for relatively long times. To the contrary, we show that the method works well in both localized and delocalized phases, and makes reliable predictions for a number of quantities including infinite-temperature autocorrelation functions. We complement our findings with rigorous incremental bounds on the truncation error. This approach shows that in practice, the exploration of intermediate-scale time evolution may be more feasible than is commonly assumed, challenging near-term quantum simulators.
\end{abstract}

\maketitle

Taming the exponential complexity of many-body quantum systems remains one of the biggest challenges in modern physics. Exact numerical simulations provide the gold standard in accuracy, however, the computational cost quickly becomes prohibitive above a few tens of particles, and even rapid developments in computing power cannot outpace the exponential scaling of the complexity of fully solving a many-body quantum system. While there are efficient methods able to estimate the ground states of various quantum systems captured by local Hamiltonians -- including tensor network and quantum Monte Carlo techniques -- the issue of complexity becomes even more of an obstacle for time evolution. Time evolution of a given quantum state under the action of a local Hamiltonian is \emph{BQP complete} in worst case complexity. For this reason, one cannot hope to find universal classical methods that can accurately and efficiently simulate this evolution for all time and all local Hamiltonians~\cite{PhysRevLett.100.010501}. While the ultimate  goal may be the development of flexible and reliable \emph{quantum simulators}  \cite{CiracZollerSimulation,BlochSimulation,Trotzky} able to directly realize many models of interest in the near-term we must continue to rely upon classical computers in order to simulate quantum matter.

To that end, many highly effective numerical techniques have been developed to study many-body quantum systems subject to controlled and clear approximations. Leading the charge are \emph{tensor network methods}~\cite{Schollwock11,Orus-AnnPhys-2014}, instances of variational methods that build on tensor networks, particularly \emph{matrix product state} (MPS) approaches in one dimension and \emph{projected entangled pair states}  (PEPS) techniques in two dimensions. These methods work well for ground states and short time evolution, but are limited
in the way they can capture dynamics, a state of affairs sometimes dubbed the `entanglement barrier'. This core limitation stems from
the generation of entanglement, as highly entangled systems require large bond dimensions, giving rise to computationally intractable situations. \emph{Quantum Monte Carlo techniques} 
\cite{RevModPhys.73.33} are also widely used
-- including for non-equilibrium dynamics~\cite{Carleo+12,Carleo+14} -- however, they suffer from the well-known sign problem and stability issues. Dynamical mean-field theory can also capture quantum dynamics~\cite{RevModPhys.68.13,Aoki+14}, 
but again stability matters arise.
These obstacles all reflect the computational hardness of the task.

\begin{figure}
    \centering
    \includegraphics[width=\linewidth]{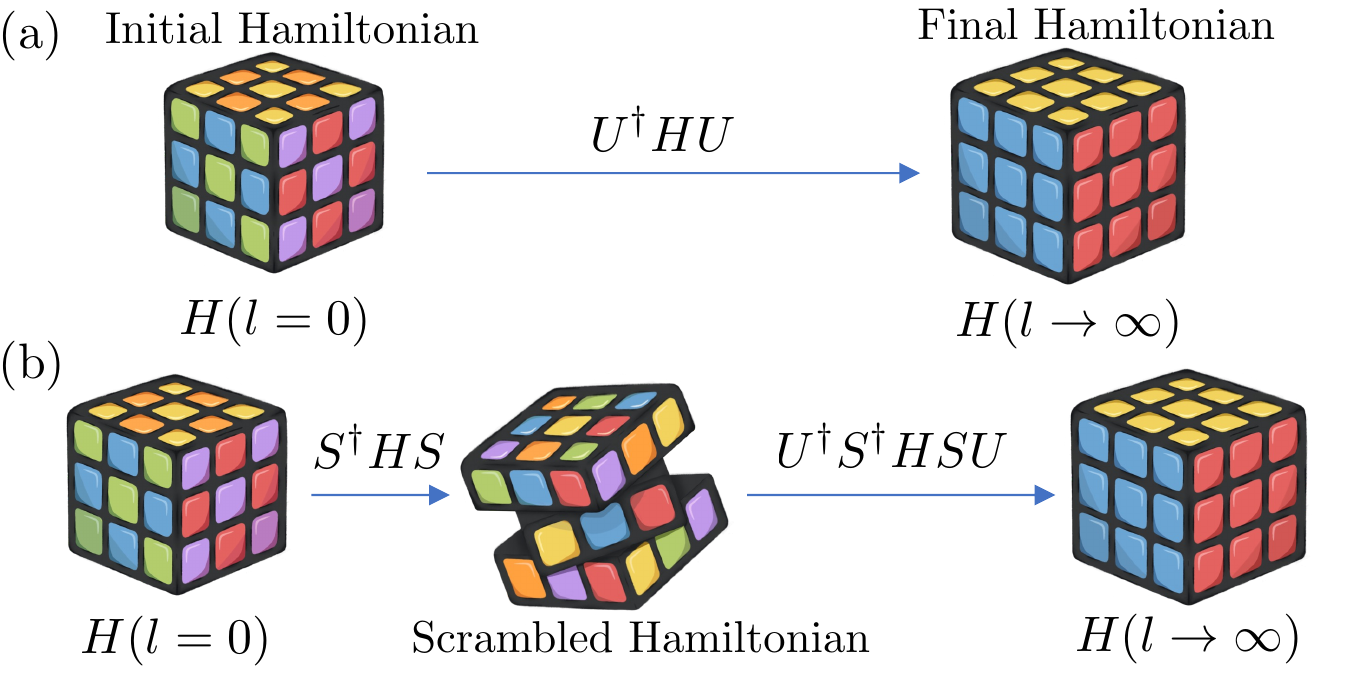}
    \caption{A cartoon illustration of the scrambling process. a) The conventional CUT process, using a single unitary transform $U$ to diagonalise a Hamiltonian $H$, smoothly transforming from the initial basis ($l=0$) to the diagonal basis ($l \to \infty$). b) The scrambling transform $S$ first induces an `effective disorder' even in completely clean systems, which allows established CUT techniques to then take over and efficiently diagonalise the `scrambled' Hamiltonian $S^{\dagger}HS$ with a second unitary transform $U$.}
    \vspace{-0.2in}
    \label{fig.scramble}
\end{figure}

In this work, we develop a radically different approach to addressing the issue of time evolution in closed quantum systems. Combining the established method of \emph{continuous unitary transforms} (CUTs) -- also known as `flow equations'~\cite{Brockett91,Chu+90,Chu94,Glazek+93,Glazek+94,Wegner94,Kehrein07,PhysRevLett.100.175702} -- with the newly developed method of \emph{scrambling transforms}, we present a flexible and powerful approach to diagonalizing large Hamiltonians and computing time evolution to very long times. The key ingredient in our work is the use of scrambling transforms to improve the convergence properties of CUT-based methods, significantly improving their accuracy and validity. We demonstrate the potential of this technique by computing the dynamics of disordered quantum systems in one and two dimensions. The limitation is a very different one compared to tensor network approaches: here, it is not entanglement that provides a limitation, but the accuracy of the approximate transform used. Heuristically, we find that in practice this restriction is less severe than overcoming the entanglement barrier.

\begin{figure}
    \centering
    \includegraphics[width=\linewidth]{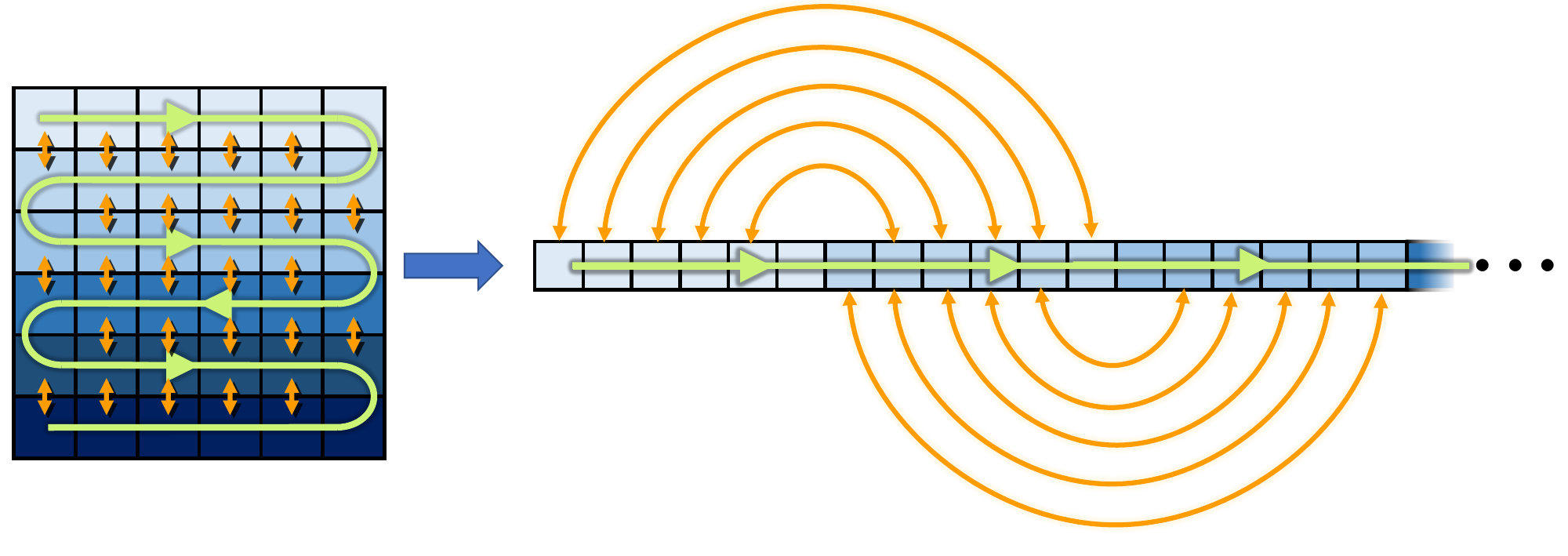}
    \caption{An illustration of how a two-dimensional lattice can be mapped onto a one-dimensional chain with correlated long-range hopping, which can be easily handled with CUT-based techniques.}
    \label{fig.snake}
\end{figure}

We will focus on a generic system of \emph{interacting fermions}, given by the Hamiltonian
\begin{align}
    H &=  \sum_{i,j\in {\cal L}} H^{(2)}_{ij} :c^{\dagger}_i c_j: + \sum_{i,j,k,q 
    \in {\cal L}
    } H^{(4)}_{ijkq} :c^{\dagger}_i c_j c^{\dagger}_k c_q:, \nn \\
    & =: H^{(2)} + H^{(4)} ,
    \label{eq.Hrun}
\end{align}
where $: ... :$ represents normal-ordering with respect to the vacuum, and \je{$|{\cal L}|=:L$ is the system size in the total number of modes}. 
We make no assumptions as to the form of the couplings, nor the dimensionality of the system: The complexity of the calculation is set by the total number of lattice sites
\je{$L$}, not by their geometry or size of the local Hilbert spaces. A two-dimensional system can be unfolded onto a one-dimensional system with long-range hopping, as sketched in Fig.~\ref{fig.snake}, which does not pose a problem for CUT-based techniques. A similar procedure can be followed in three dimensions.

Flow equation methods diagonalise the Hamiltonian by successively applying infinitesimal unitary transforms $dU(l) = \exp(-\eta(l) \ud l) = 1 - \eta(l) \ud l$, where $\eta(l)$ is the generator and $l$ represents a fictitious ``flow time'' such that $l=0$ is the initial Hamiltonian, and the parametrized Hamiltonian $H(l):=U^{\dagger}(l) H U(l)$ becomes diagonal in the limit $l \to \infty$, where the full unitary transform $U(l) = \mathcal{T}_l \exp( -\int_0^l \eta(l') \ud l')$ is a time-ordered integral over flow time $l$. The diagonalization procedure can be recast as solving the equation of motion $\ud H/ \ud l = [\eta(l),H(l)]$~\cite{Wegner94,Kehrein07}. We store $H^{(2)}$ as a matrix \je{with $O(L^2)$ entries}
and $H^{(4)}$ as a
tensor of \je{order four with $O(L^4)$ real entries}, and employ a similar procedure for the generator $\eta(l) =: \eta^{(2)}(l) + \eta^{(4)}(l)$. This allows the relevant commutators to be computed efficiently as the sum of all one-point contractions of pairs of matrices \je{or} tensors~\cite{Thomson+23}. The main consequence of fermionic statistics is the minus signs which arise when computing these contractions; the method can be applied to bosons with minor changes.

\begin{figure}[t!]
    \centering
    \includegraphics[width=\linewidth]{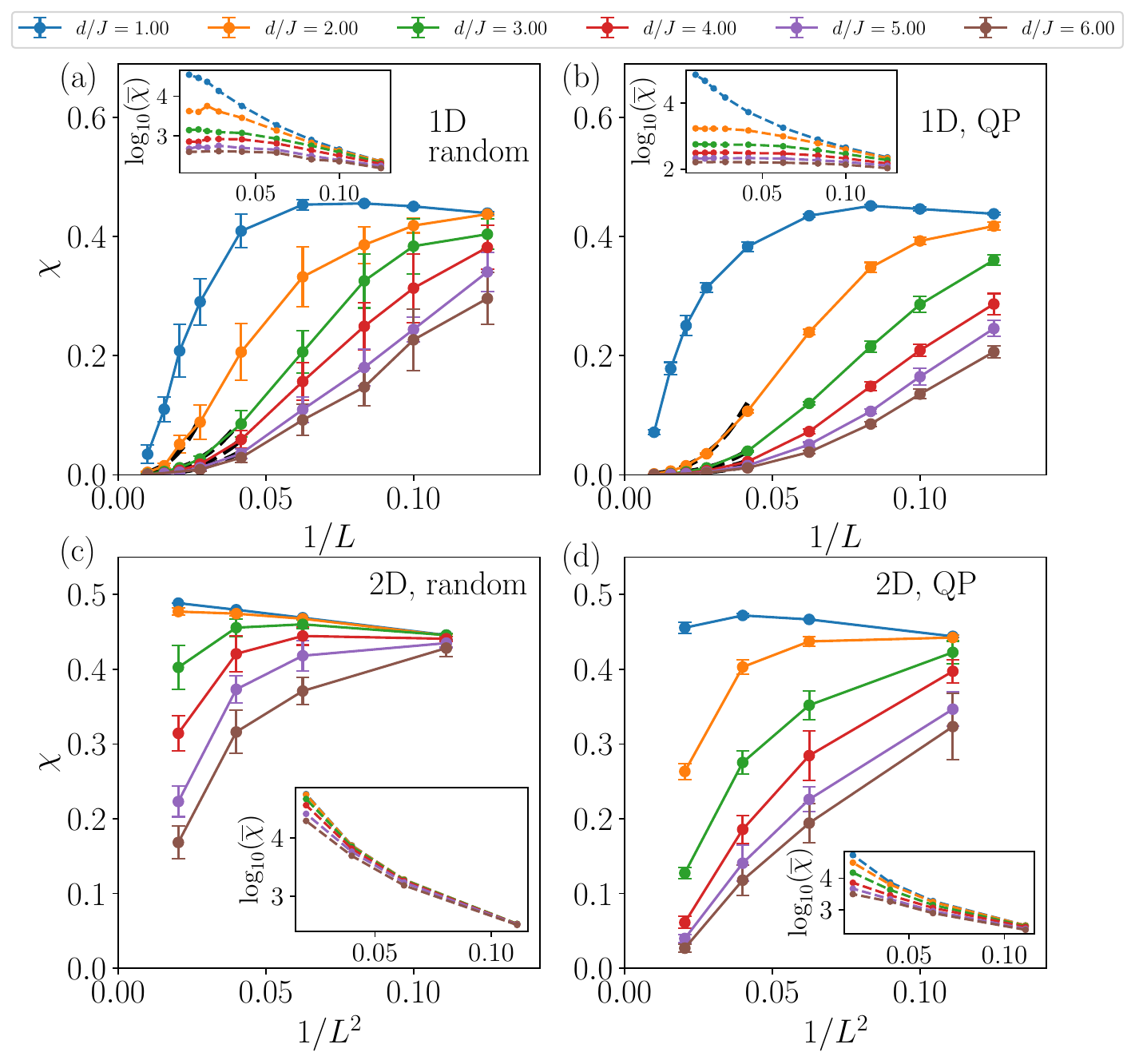}
    \caption{The complexity $\chi$ of the transformed creation operator $c^{\dagger}_i$ in the middle of the system. Error bars show the standard deviation over disorder realizations. Panels (a) and (b) show the results in one dimension ($L=8,10,12,16,24,36,48,64,100$) for random and quasi-periodic potentials respectively, while panels (c) and (d) show the same in two dimensions ($L^2 = 9,16,25,36,49,64,100$). Dashed black lines close to the origin in panels (a) and (b) are fits with the form $\chi \propto 1/L^3$, valid for large systems and strong disorder. Insets show the un-normalised complexity (i.e., the numerator of Eq.~(\ref{eq.chi})), which tends to a constant in strongly-disordered one-dimensional chains, but grows in two dimensions even for strong disorder.}
    \label{fig.complexity}
\end{figure}

A common choice of generator is $\eta(l) := [H_0,V(l)]$, where $H_0(l)$ and $V(l)$ are, respectively, the diagonal and off-diagonal parts of the Hamiltonian. 
\je{In what follows, we will make use of the symbol $V$
whenever referring to off-diagonal elements.}
This is often known as the \emph{Wegner generator}~\cite{Wegner94,Kehrein07}. 
The diagonalisation can be seen from the fact that the squared 
$\| V(l)\|_2^2$
is non-increasing in the fictitious time $l$ as
$\ud \| V(l)\|_2^2 / \ud l = -2 \|  \eta(l)\|_2^2 \leq 0$ (see, e.g., Ref.~\cite{Monthus16}).
Convergence relies upon the model in question having a clear separation of energy scales in the initial basis. 
Models where this is not true -- such as homogeneous systems, or disordered systems with many near-degeneracies (known as resonances) -- cannot be fully diagonalised by this generator, as they act like unstable fixed points. Perturbing the Hamiltonian away from this fixed point can allow the flow to begin, however, small perturbations can result in long convergence times while large perturbations improve convergence but risk changing the underlying physics. Here, we resolve this by introducing \emph{scrambling transforms}, which are targeted unitary transforms aimed at lifting degeneracies which the Wegner procedure alone is unable to resolve. As they are unitary, they cannot change the underlying physics: they simply act to `prepare' the Hamiltonian in a basis more amenable to being diagonalized by the conventional Wegner flow. The procedure is sketched in Fig.~\ref{fig.scramble}.
The (infinitesimal)
scrambling transform takes the form $dS(l)=\exp(-\lambda(l) \ud l)$, with a generator $\lambda(l)$ given by
\begin{align}
\lambda_{ij}(l) :=
\begin{cases}
      \textrm{sgn}(i-j) J_{ij}(l) :c^{\dagger}_i c_j: & \text{if}\ J_{ij}(l) \geq \delta h  \\
      0 & \text{otherwise,}
    \end{cases}
    \label{eq.scramble}
\end{align}
with $\delta h = \varepsilon |h_i(l) - h_j(l)|$, where $\varepsilon>0$ is the threshold parameter which controls how easily the scrambling transform triggers. For $\varepsilon =0$, this reduces to the \emph{Toda-Mielke generator}~\cite{Mielke98,Monthus16}. Here, we use $\varepsilon=0.5$. The full scrambling transform $S(l)$ can be written as a time-ordered integral over $dS(l)$. It is employed at the beginning of the flow, and during the diagonalization procedure if degeneracies are encountered.

\begin{figure*}
    \centering
    \includegraphics[width=0.85\linewidth]{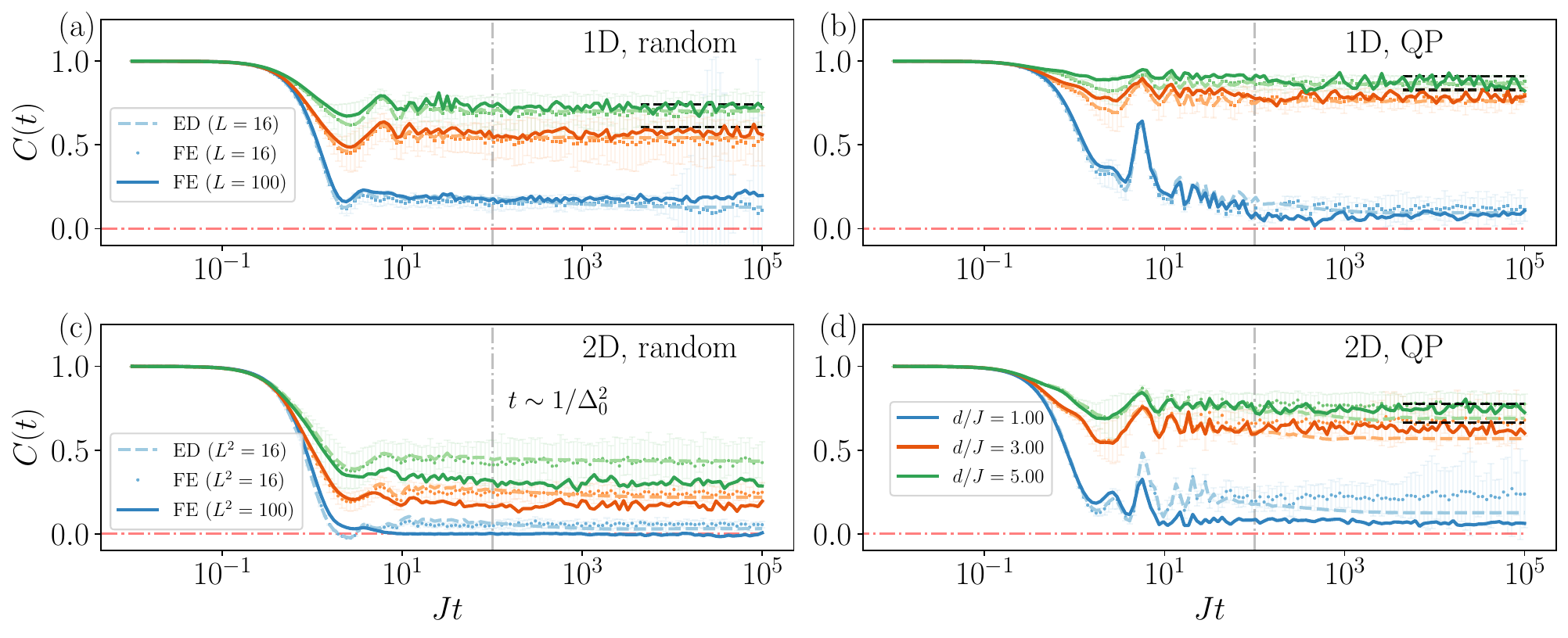}
    \caption{Infinite-temperature correlation functions shown for a variety of disorder types, strengths, dimensionalities and system sizes. The grey dot-dashed vertical line indicates the approximate timescale beyond which accuracy cannot be guaranteed, however, the results typically remain reasonable until much longer timescales. In all plots, the result from \emph{exact diagonalisation} (ED) is shown as a pale dashed line. The equivalent result from the \emph{flow equation} (FE) method is shown by circular markers, and the result obtained using flow equations for a much larger system is indicated by a dark solid line. Black dashed lines show the long-time average computed directly \emph{without} explicit time evolution, valid at strong (quasi)disorder only. The results are averaged over $N_s \in [64,256]$ disorder realizations, depending on system size. Error bars indicate the variance over disorder realizations, for clarity shown only for the flow equation results for $L=16$ ($L^2 = 4 \times 4$ in two dimensions). In both $D=1$ and $D=2$, the quasi-periodic potential exhibits most much robust localization at large values of $d/J$, but by contrast also exhibits more complete thermalisation at low values of $d/J$ due to the underlying single-particle phase transition at $d/J=2$.}
    \label{fig.itc}
\end{figure*}

The particular scrambling transform used here is quadratic and does not induce any new higher order terms, however, the action of the Wegner generator will typically lead to the generation of new terms containing six or more fermionic operators, similar to the way that such terms can arise in renormalization group procedures. The central approximation of the CUT technique is that the Hamiltonian must be truncated, and terms above a certain order neglected. We shall present rigorous error bounds later; for the moment, we emphasize that in cases where the method is insufficiently exact, higher order terms can be systematically included until the desired precision is reached, at a cost polynomial in the system size.

We will investigate initial local Hamiltonians of the form $H = \sum_{i\in {\cal L}} 
[ h_i :n_i: + J (:c^{\dagger}_i c_{i+1}: + H.c.) + \Delta_0 :n_i n_{i+1}:]$, using open boundary conditions, with $J=1$ and $\Delta_0 = 0.1$. In one dimension, this Hamiltonian maps onto the XXZ chain via a Jordan-Wigner transform. We diagonalise these Hamiltonians in both one and two dimensions, for two different choices of $h_i$: random disorder ($h_i \in [-d,d]$) and \emph{quasi-periodic} (QP) potentials (in
\je{one spatial dimension} $h_i := d \cos(2 \pi i/\phi + \theta)$, with $\phi := (1 + \sqrt{5})/2$ and $\theta$ a (real) randomly chosen phase that plays the role of a `disorder realization'; in 
\je{two dimensions}
given by $h_{i} := d (\cos(2 \pi i_x/\phi + \theta) + \cos(2 \pi i_y/\phi_2 + \theta_2))$, where $(i_x,i_y)$ represent the coordinates of lattice site $i$, $\phi_2 = 1 + \sqrt{2}$ and the $\theta_2$ is another random phase). For simplicity, we shall refer to $d$ as the `disorder strength' in both cases. 
The end point is an
\je{(approximately)}
diagonal Hamiltonian
\begin{align}
    \tilde{\je{H}} &= \sum_{i\in {\cal L}} \tilde{h}_i :\tilde{n}_i: + \sum_{{i,j}\in {\cal L}} \Delta_{ij} :\tilde{n}_i \tilde{n}_j: +\mathcal{R},
\end{align}
where $\mathcal{R}$ 
represents neglected higher-order terms, typically of order $\je{O}(\Delta_0^2)$ and higher. The interaction coefficients have been shown to decay exponentially with distance in (quasi)disordered systems, $\Delta_{ij} \propto \textrm{e}^{-|i-j|/\xi}$~\cite{Rademaker+16,Rademaker+17,Thomson+18,Thomson+20b,Thomson+20c,Thomson+23}.

\begin{figure*}[ht]
    \centering
    \includegraphics[width=.9\linewidth]{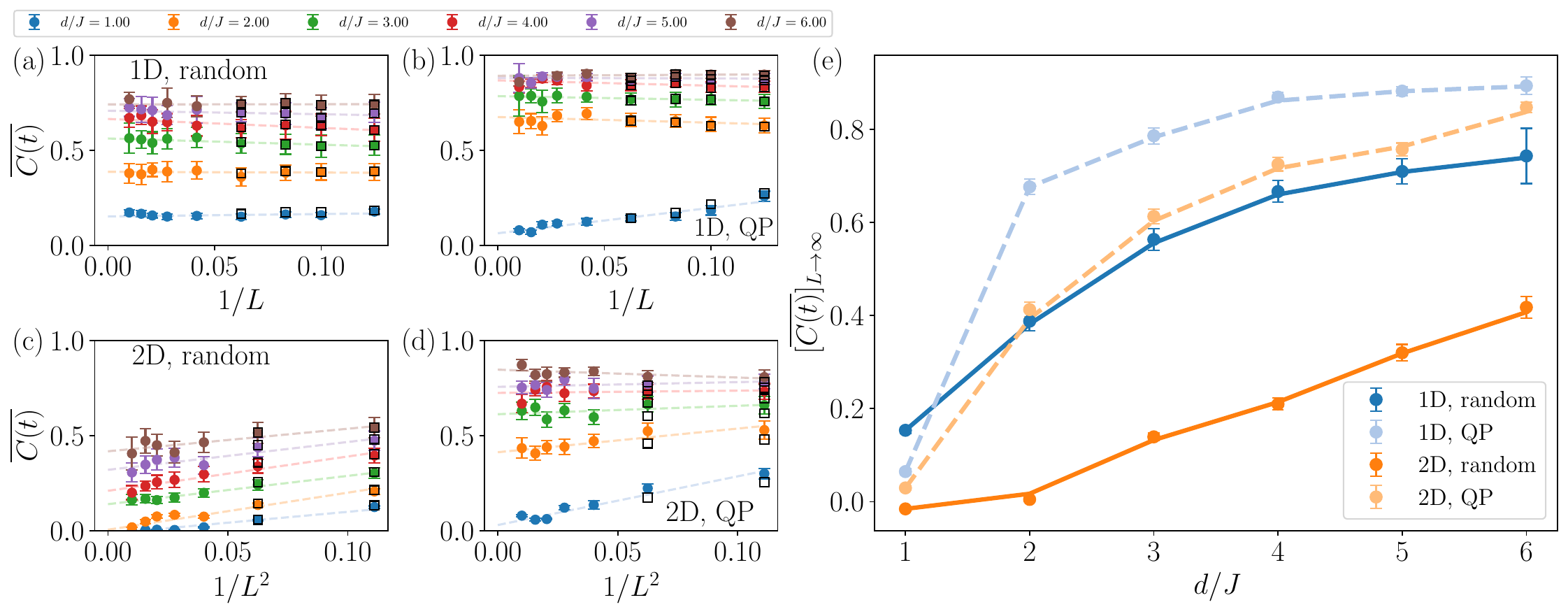}
    \caption{Finite-size scaling results for the long-time average of the infinite-temperature correlation function $\overline{C(t)}$, averaged over a time window of $Jt \in [50,10^3]$, and then averaged over $N_s \in [32,1024]$ disorder realizations depending on system size. (a-d) Results for various system sizes, disorder types and strengths, and dimensionalities. Error bars indicate the variance over disorder realizations. The black squares indicate the results obtained from exact diagonalisation, shown for small system sizes only. System sizes are $L=8,10,12,16,24,36,48,64$ and $100$ in 
    \je{one dimension}, and $L^2 = 9,16,25,36,49,64$ and $100$ in two dimensions. (e) The result of linearly extrapolating to $L \to \infty$. Error bars indicate the uncertainty in the fits [shown as dashed lines in panels (a-d)] to extract the scaling behavior; lines are smoothed guides to the eye.}
    \label{fig.fss}
\end{figure*}

Once the Hamiltonian has been diagonalized, it is possible to obtain a closed-form solution (within a given truncation scheme) to the Heisenberg equation of motion for any operator $O$ expressed in the diagonal basis. The operator must first be transformed according to the flow equation $\ud O/ \ud l = [\overline{\eta}(l),O(l)]$, where $\overline{\eta}(l)$ collectively denotes both the scrambling and Wegner generators. This transformed operator also contains valuable information about the locality of the unitary transform, and can be used to extract both a localization length and as a measure of the ``complexity'' of the diagonalization procedure, 
which can
be linked to the existence of Lieb-Robinson bounds in flow time~\cite{Hastings22}. Specifically, the transformed creation operator takes the form $c^{\dagger}_i = \sum_j A^{(i)}_j \tilde{c}^{\dagger}_j + \sum_{j,k,q} B^{(i)}_{jkq} \tilde{c}^{\dagger}_j \tilde{c}^{\dagger}_k \tilde{c}_q$, and higher-order terms are neglected.
A measure of the complexity of the transformed operators is given by the fraction of non-zero terms which appear in this operator expansion. In practice, we choose a cutoff value $\epsilon = 10^{-6}$ below which we consider terms to be zero. The \emph{complexity} 
in this sense is defined as
\begin{align}
    \chi(\epsilon) = \frac{| \{ x \in (A \cup B) | x^2 > \epsilon^2 \} |}{| \{ x \in (A \cup B) \} |},
    \label{eq.chi}
\end{align}
where $(A \cup B)$ represents the set of all coefficients $A_i$ and $B_{ijk}$ in the operator expansion of $c^{\dagger}_i$. We also define $\overline{\chi(\epsilon)} = | \{ x \in (A \cup B) | x^2 > \epsilon^2 \} |$. Results are shown in Fig.~\ref{fig.complexity}, demonstrating a qualitative difference between one and two spatial dimensions. In one dimension, we find a phase where $\overline{\chi(\epsilon)}$ tends to a constant, and $\chi(\epsilon) \propto (1/L)^3$ for large system sizes, indicating a `low complexity'
situation
at strong disorder, as well as a higher complexity phase at small values of $d$ where $\overline{\chi(\epsilon)}$ increases rapidly with system size, suggestive of thermalisation. In two dimensions, we find that $\overline{\chi(\epsilon)}$ always increases, although in the case of the QP potential at large values of $d$, it increases sufficiently slowly that the normalised complexity $\chi(\epsilon)$ still vanishes. By contrast, for small values of $d$ in two dimensions, the complexity $\chi(\epsilon)$ remains much larger than zero for all system sizes studied here. This suggests a slow crossover from a high complexity phase -- consistent with the expectation of thermalisation at small values of $d$ -- to a low complexity phase with anomalous thermalisation properties.

Previous work which have used CUT methods to compute non-equilibrium dynamics~\cite{Thomson+18,Thomson+20,Thomson+20b} employed a computationally costly inversion of the unitary transform in order to obtain time-evolved operators in the original basis. Here, we circumvent this limitation and directly obtain the infinite-temperature autocorrelation function, a highly non-trivial and challenging quantity to compute. The thermal expectation value of any arbitrary operator $O$ is given by $\langle O \rangle = \textrm{Tr}[\exp(-\beta \je{H}) O]/\textrm{Tr}[\exp(-\beta \je{H})]$, where $\beta =1/T$ is the inverse temperature (in units of $k_B=1$). In the limit of $T \to \infty$, the expectation value becomes 
a uniform average over eigenstates, which in the diagonal basis are trivial product states. We approximate this average for large systems by randomly sampling $\mathcal{N}_s \in [50,256]$ half-filled eigenstates, with an associated standard error that scales 
\je{as}
$1/\sqrt{\mathcal{N}_s}$. Specifically, we compute the dynamical autocorrelation function
\begin{align}
    C(t) = 4 \langle (n_i(t) - 1/2)(n_i(0) - 1/2) \rangle.
    \label{eq.fermion_corr}
\end{align}
To minimize boundary effects, we choose $i$ to be in the center of the system. We benchmark the performance of this approximation by comparison with exact diagonalization, making use of dynamical quantum typicality~\cite{Goldstein+06,Reimann07,Bartsch+09,Bartsch+11,Elsayed+13,Richter+19,Heitmann+20,Chiaracane+21} to compute the infinite temperature correlation function. This is a highly demanding quantity that can be extremely challenging to compute with other methods, but can be obtained very 
efficiently 
with the flow equation approach. Results for system size $L=16$ ($L^2=4\times 4$ for two dimensions) are shown in Fig.~\ref{fig.itc}, demonstrating excellent agreement with exact diagonalization in both one and two dimensions, even to timescales as long as $tJ = 10^5$. This is beyond the naive expectation that the accuracy should break down beyond timescales $tJ \sim 1/\Delta_0^2$, corresponding to the typical inverse magnitude of the terms cut off by the truncation. We also show results for larger system sizes, demonstrating that they remain smooth and under control. Fig.~\ref{fig.fss} shows results for system sizes up to $L=100$ ($L^2= 10 \times 10$ in two dimensions), along with a linear fit indicating the $L \to \infty$ behavior. Strikingly, very little dependence on system size is observed in \je{one dimension}, although in two dimensions there is a slow trend towards decreasing values of $C(t)$ as the system size increases, except for strong quasiperiodic potentials. This is consistent with the expectation that many-body localization may be ultimately unstable in two dimensions, although these results suggest one must go to very large system sizes and long times to see signs of this potential instability. For weak random disorder in \je{two dimensions}, the linear fit for large values of $L$ breaks down, suggesting our results are likely to overestimate the long-time value of $C(t)$ as $L \to \infty$.  
\je{Additionally},
at any finite order of truncation there may still exist higher-order processes which could contribute to thermalisation on very long timescales. Nonetheless, for a given truncation scheme we can make precise statements about the validity of this technique.

To do so, we develop an incremental bound on the error in the unitary transform. If at each flow time step we discard all newly-generated terms above fourth-order, we obtain
\begin{align}
    H(l+\ud l) &= H(l) + \ud l [\eta(l),H(l)] \nn \\
    & =: H(l) + \ud l (dH(l) + A(l)),
\end{align}
where $dH(l) = [\eta^{(2)}(l),H^{(2)}(l)] +  [\eta^{(2)}(l),H^{(4)}(l)] +  [\eta^{(4)}(l),H^{(2)}(l)]$ represents the terms of the flow which are kept and $A(l) =  [\eta^{(4)}(l),H^{(4)}(l)]
+{\cal T}$ represents the 
truncation error, where the higher-order terms ${\cal T}$ are assumed to be negligible in what follows. The norm of the truncation error $A(l)$ at each infinitesimal time step is 
\je{upper bounded}
by
\begin{align}
    \|A(l)\|_2 &= \| [\eta^{(4)}(l),H^{(4)}(l)] \|_2 \leq \sqrt{2} \|\eta^{(4)}(l)\|_2 \|H^{(4)}(l)\|_2 \nn \\
    &\leq 2 \|H^{(4)}_0(l)\|_2 \|V^{(2)}(l)\|_2 \|H^{(4)}(l)\|_2 
    \label{eq.incremental_err}
\end{align}
using the sub-multiplicativity of the $\|.\|_2$ norm~\cite{Bottcher+08}.
The total truncation error in flow time can be written as an integral of Eq.~(\ref{eq.incremental_err}) 
\begin{align}
    \varepsilon_T
    \leq 2 \int_{0}^{l_{\textrm{max}}} \ud l  \|H^{(4)}_0(l)\|_2 \|V^{(2)}(l)\|_2 \|H^{(4)}(l)\|_2
    \label{eqtruncation_err}
\end{align}
over $l$. 
Typical values of $V^{(2)}(l)$ decay exponentially in flow time, i.e., $[V^{(2)}(l)]_{ij} \propto \exp(-(h_i-h_j)^2 l) [V^{(2)}(0)]_{ij}$. Assuming random disorder drawn from a box distribution of width $[-d,d]$, such that the mean value of this squared energy difference is $2d^2/3$, and that the largest parts of the interaction tensor remain proportional to the initial interaction strength (as new terms induced by the flow should always be smaller than the initial interactions), the error can be approximated as
\begin{align}
    \varepsilon_T & \propto J_0 \Delta_0^2 \int_{0}^{l_{\textrm{max}}} \ud l \phantom{.} \textrm{e}^{-l 2d^2/3}  \underrel{l_{\textrm{max}} \to \infty}{=} \frac{3}{2}\frac{J \Delta_0^2}{d^2}.
\end{align}
In the case of weak (quasi)disorder, the disorder bandwidth $d$ is replaced by the effective bandwidth $\tilde{d} \geq d$ induced by the scrambling transform~\cite{SM}. A numerical analog can be computed by replacing the Hilbert-Schmidt (\je{or Frobenius}) 
norms in Eq.~(\ref{eq.incremental_err}) 
with tensor Frobenius norms; the typical truncation error at each flow time step is well below one percent~\cite{SM}.

The above analysis indicates that energy differences below $\je{O}(J \Delta_0^2/d^2)$ cannot be reliably resolved, implying that the method will break down on timescales on the order $t \propto d^2/(J \Delta_0^2)$ when oscillations at corresponding frequencies $\omega \sim \je{O}(J \Delta_0^2/d^2)$ become relevant for the dynamics. The accuracy of the method can be systematically improved by incorporating additional higher-order terms into the truncated Hamiltonian, allowing accurate simulations of quantum dynamics to even longer times (proportional to $1/\Delta_0^3$ at the next order of approximation) with a computational cost that remains polynomial in the system size. Future developments in massively parallel implementations of the \emph{tensor flow equation method}~\cite{Thomson+23} used in this work, as well as advances in computer hardware, will facilitate extension of this method to larger system sizes, longer timescales, stronger interactions, and additional physical systems (including both driven~\cite{Thomson+20c,PyFlow} and dissipative~\cite{Rosso+20} systems, which have been previously studied with CUT-based techniques). Scrambling transforms may be of interest in a variety of other contexts, as they are essentially a way of transforming a highly entangled system into a simpler representation that is easier to simulate. 

\je{We end the discussion by briefly comparing the findings with
\emph{tensor network methods}. Standard tensor network methods are challenged in time evolution by the
exponentially growing bond dimension that is required to accommodate the states faithfully
in time, for the linear growth of entanglement. 
Some steps have already been taken to allow tensor networks to access longer times~\cite{LongTimes,Krumnow2016,Folding,PollmannMPSLongTimes}, e.g.,
by means of \emph{folding techniques}~\cite{Folding}
or adaptive \emph{mode transformations}~\cite{Krumnow2016}: The latter overcomes the prejudice that
quantum states have to be represented in real space: One can co-rotate the frame of mode transformations,
so that only the entanglement between these effective modes need to be accommodated.
The ideas introduced here show that one can go further than that, by overcoming the prejudice that
the fermionic mode transformation needs to be linear: accepting small truncation errors in the procedure,
one can even fully diagonalise the Hamiltonian to a good approximation. There are good reasons to believe that this is a favourable mindset to reach long simulation times. First connections between flow equation and tensor network 
methods have been made \cite{Flow}, 
anticipating the \emph{time-dependent variational
principle} based on a differential geometric picture~\cite{2011PhRvL.107g0601H}. It is
conceivable that the ideas introduced here
can be further merged with tensor network
techniques, as one could think of
final Hamiltonians that are not treated
as fully diagonal ones.
There is also the intriguing possibility of combining scrambling transforms and other CUT-based techniques with tensor network approaches, such as entanglement-based continuous unitary transforms~\cite{Sahin+17}, which may allow tensor network methods to break through the entanglement barrier and access much longer timescales than are currently available.}

\je{In this work, we have introduced a flow-based method equipped with scrambling transforms
that allow for simulating interacting fermionic quantum many-body systems to good accuracy for intermediate and long times.}
Adding modern machine learning techniques into the mix could also allow for the development of efficient \emph{data-driven} scrambling transforms tailored for specific problems. Such a classical development can also be seen as a challenge
to \emph{dynamical quantum simulators} \cite{CiracZollerSimulation,Trotzky} which aim to
probe non-equilibrium properties of quantum matter beyond the reach of classical computers. These are exciting avenues for future progress.

\section*{Acknowledgements}

This project has received funding from the European Union's Horizon 2020 research and innovation programme under the Marie Skłodowska-Curie grant agreement No.101031489 (\href{http://ebqm.info}{Ergodicity Breaking in Quantum Matter}) and under the Quantum Flagship (PASQuanS2,
Millenion), as well as support from the NVIDIA Corporation through the Academic Hardware Grant Program, the BMBF (\je{FermiQP}, MuniQC-Atoms), and the DFG (CRC 183). \je{S.~J.~T.~thanks} J.~Catton for providing the artwork used in Fig.~\ref{fig.scramble}. Data and code are available at Refs.~\cite{PyFlow,data}.

\section*{Methods}

This work made use of the \texttt{PyFlow} library~\cite{PyFlow}, developed and maintained by S.~J.~T.~, based on Ref.~\cite{Thomson+23}.

\subsection{Computing and integrating the flow equation}

All commutators computed in this work follow the scheme of Ref.~\cite{Thomson+23}, where the representation of the Hamiltonian in terms of a quadratic component (stored in memory as a matrix) and quartic component (stored as a tensor) allow the commutators to be recast in terms of matrix/tensor contractions, which are highly-optimised linear algebra operations that can be performed efficiently on modern computing hardware. A complete description is contained in the Supplemental Information~\cite{SM}. We use vacuum normal-ordering, such that higher-order terms in the running Hamiltonian have no feedback onto lower-order terms. The incorporation of additional non-perturbative corrections due to different choices of normal-ordering has previously been done in the time-independent scenario~\cite{Thomson23}, but in the non-equilibrium setting is left for future work. This would require specifying a particular reference state with respect to which the corrections are computed, and depending on the structure of this state, the resulting corrections may be hard to vectorize, complicating their implementation on \emph{graphics processing units} (GPUs). Calculations for all system sizes with more than a total of $16$ lattice sites were performed on GPUs (specifically, NVIDIA RTX A5000 GPUs with 24Gb RAM and NVIDIA RTX 2080Ti GPUs with 12Gb RAM) using single precision arithmetic. 

The flow equation $\ud H
\je{l}/ \ud l = [\overline{\eta}(l),H(l)]$ is solved using a mixed 4th/5th order Runge-Kutta integration method as implemented in the \texttt{JAX} library~\cite{jax2018github}, making use of an adaptive stepsize algorithm for high accuracy. The maximum integration time used was $l_{\rm max}=1000$, and the integration is stopped before then if the Hamiltonian is diagonalized to the target accuracy, which we choose to be when $\textrm{max}[|V^{(2)}|]<10^{-6}$ and $\textrm{max}[|V^{(4)}|]<10^{-3}$. Results for longer integration times showed no significant increase in accuracy, despite incurring a significantly higher computational cost. This is because the running Hamiltonian $H(l)$ approaches full diagonalisation only asymptotically at large values of $l$, so the use of larger values of $l_{\rm max}$ leads to diminishing returns.

\subsection{Computing the dynamics}

The transformed number operator is reconstructed from the transformed creation/annihilation operators 
\je{for large fictitious time}
\begin{equation}
n_i(l \to \infty) = c^{\dagger}_i(l \to \infty) \times c_i(l\to \infty), 
\end{equation}
with the creation operator given by 
transformed creation/annihilation operators 
\begin{equation}
c^{\dagger}_i(l \to \infty) = \sum_j A^{(i)}_j \tilde{c}^{\dagger}_j + \sum_{j,k,q} B^{(i)}_{jkq} \tilde{c}^{\dagger}_j \tilde{c}^{\dagger}_k \tilde{c}_q 
\end{equation}
and the annihilation operator obtained by taking its Hermitian conjugate $c_i(l \to \infty) = (c^{\dagger}_i(l \to \infty))^{\dagger}$. Multiplying these together allows us to reconstruct the number operator including terms up to sixth-order in the fermionic creation/annihilation operators for the diagonal basis, $\tilde{c}^{\dagger}_i$ and $\tilde{c}_i$. The number operator can then be time-evolved in the diagonal basis according to the Heisenberg equation of motion, neglecting newly-generated higher-order terms, resulting in a closed-form solution. This step is performed on CPUs rather than GPUs due to memory limitations, and is a prime candidate for future efficiency improvements. At long times, near-degenerate single-particle eigenvalues can still lead to divergent terms in this solution (consistent with the expectation that the simulation of a BQP-hard problem will eventually run into accuracy issues on a classical computer), however, these terms are strongly suppressed, arising only very rarely and at very long times. To avoid these rare scenarios dominating the averaged data, in the \je{two dimensional} dynamical data we exclude disorder realizations where the maximum value of $|C(t)|>1.1$. (Alternatively, we could have used the typical rather than mean value of $C(t)$.) See the Supplementary Information for full details of the calculation and where divergent terms arise from. In \je{one dimension}, this procedure is not necessary as the divergent terms are rare enough to have essentially no effect, as can be seen from the data in Fig.~\ref{fig.itc}. It is possible to subtract the divergent terms when they are encountered, however, we do not employ this procedure here. The long-time average can be obtained directly by setting all off-diagonal terms in the transformed number operator to zero (as when time-evolved, they acquire oscillating phases which average to zero). For systems with greater than $36$ lattice sites in total, we neglect the sixth order contributions and keep only the quadratic and quartic terms when computing the dynamics. For the systems considered here, the sixth-order terms have a negligible effect, which can be seen from the qualitative agreement between small and larger system sizes.

\subsection{Rescaling the correlation function}

As the norm of the number operator $n_i$ is not precisely conserved by the unitary transform, we rescale the correlation function for each disorder realization according to the ansatz $C(t) \je{\mapsto} c_1(C(t)-c_2)$, where $c_1$ and $c_2$ are determined by minimizing the error with respect to the short-time dynamics of the non-interacting system (as many-body interactions are essentially irrelevant at very short times). This is computationally efficient, as we get the exact dynamics of the non-interacting system essentially for free in this formalism by just retaining the quadratic components of the Hamiltonian and relevant observables. The rescaling employed in this work is justified \emph{a posteriori} by the clear agreement between the rescaled $C(t)$ and the exact result, computed for system sizes small enough for the comparison to be practical. For small enough systems, an alternative would be to construct the operator as a matrix in the full Hilbert space and renormalize it by hand, however, this is not practical for systems as large as those considered here. We emphasize  that the norm is preserved to high accuracy for sufficiently strong disorder, and the effects of this rescaling are most important for the weakly disordered systems. This is independent of any error introduced in the eigenvalues, and reflects the difficulty in simultaneously preserving the unitary evolution of both the Hamiltonian and the number operator within the same truncation scheme. The norm of the operator could in principle be exactly preserved by constructing the unitary transforms subject to additional constraints~\cite{Kehrein07}, however, in practice this is challenging to implement. This underscores the need for further work in developing more flexible generators for the types of continuous unitary transform developed here, perhaps in concert with machine learning approaches to design data-driven generators tailored for specific problems subject to 
specific hard-to-satisfy constraints.

\bibliography{BigReferences67.bib,refs.bib}

\clearpage
\onecolumngrid

\section*{Supplementary Information}

In this Supplementary Information, we provide a detailed description of the numerical procedure, additional error analysis to demonstrate the reliability of the method, and further details on the evolution of local operators under the action of a continuous unitary transform, and how time evolved operators are computed and measured using this procedure.
A \href{https://github.com/sjt48/PyFlow/blob/main/tutorial/TensorFlowEquations_StepByStepGuide.ipynb}{pedagogical tutorial to the numerical method is available in the form of an interactive Jupyter notebook}.

\subsection*{Scrambling transform}

\begin{figure}
    \centering\includegraphics[width=0.55\linewidth]{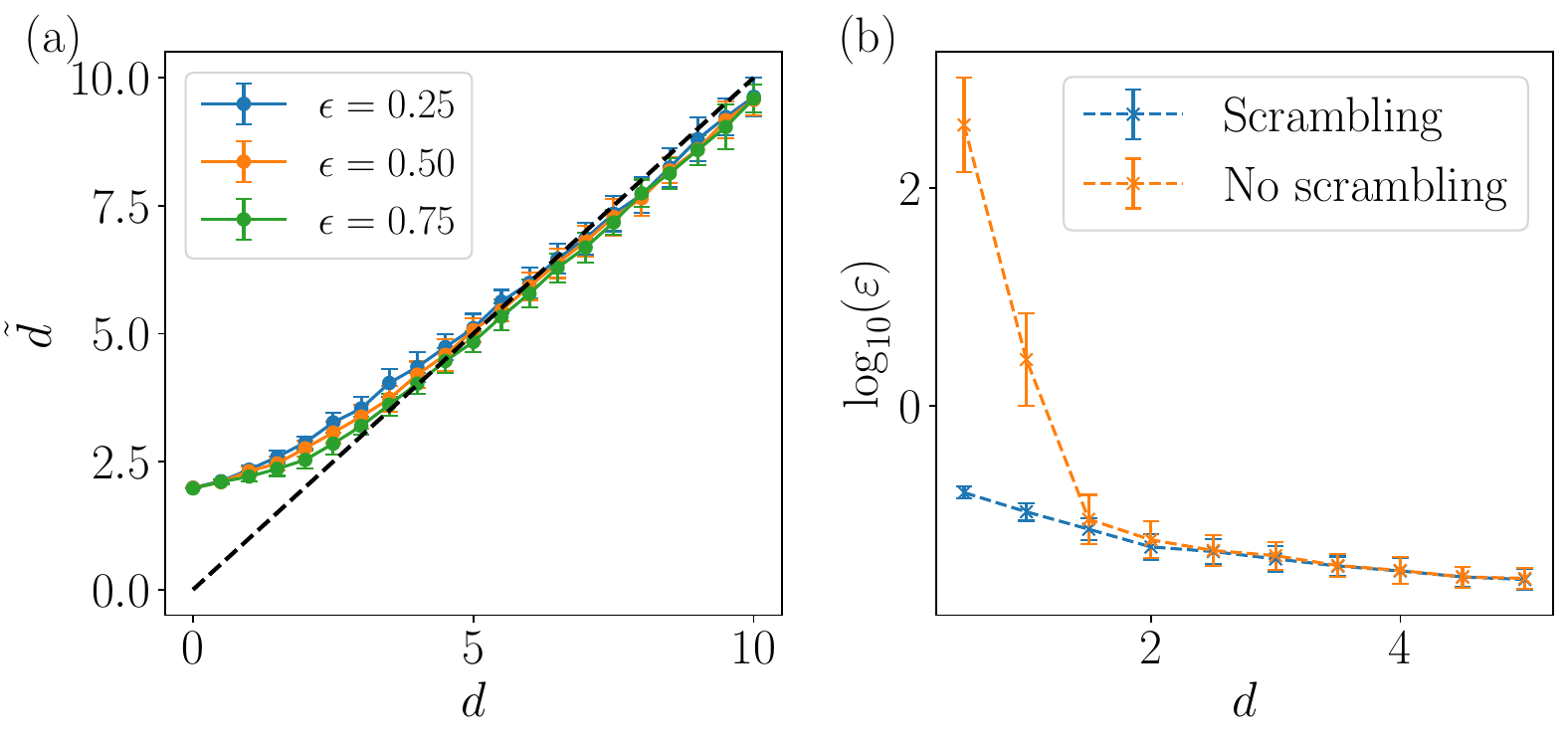}
    \caption{a) The \emph{induced} disorder bandwidth following the application of the scrambling step, with hopping $J=1$ and different values of the cutoff $\epsilon$ and the microscopic (random) disorder strength $d$. System size if $L=36$, and the results are averaged over $50$ disorder realisations. Error bars indicate the standard deviation over disorder realisations. b) A comparison of the median relative error $\varepsilon$ in the eigenvalues for the method with (blue) and without (orange) the scrambling transform, for an interacting system of size $L=10$ with $J=0.5$, interaction strength $\Delta_0 = 1.0$, maximum flow time $l_{\textrm{max}}=50$ and averaged over $192$ disorder realisations. Error bars indicate the median absolute deviation. }
    \label{fig.scramble2}
\end{figure}

Much of the main results rely on the use of \emph{scrambling transforms}, which are targeted unitary transforms used to remove degeneracies which are problematic for the Wegner generator. The scrambling transform does not have to diagonalise the Hamiltonian, even partially, but instead is used to induce an effective artificial disorder $\tilde{d} \geq d$.
The generator for the scrambling transform is 
\je{defined to be}
\begin{align}
\lambda_{ij}(l) =
\begin{cases}
      \textrm{sgn}(i-j) J_{ij}(l) :c^{\dagger}_i c_j: & \text{if}\ J_{ij}(l) \geq \delta h,  \\
      0 & \text{otherwise,}
    \end{cases}
    \label{eq.scramble2}
\end{align}
with $\delta h = \varepsilon |h_i(l) - h_j(l)|$, where \je{$\varepsilon>0$} controls how easily it triggers. In Fig.~\ref{fig.scramble2}a), we show the induced disorder bandwidth $\tilde{d} = |\textrm{max}(H_0(l^*)) - \textrm{min}(H_0(l^*))|/2$ (where $l^*$ denotes the flow time at which the scrambling transform is finished) for three different choices of cutoff $\epsilon$, computed for a non-interacting system of size $L=36$ and averaged over $50$ disorder realisations. At strong values of the microscopic disorder $d$, the scrambling transform does essentially nothing. At small values of $d$, however, the scrambling transform has a much more dramatic effect, leading to a substantial induced disorder. This implicitly disordered model can then be efficiently diagonalised by the canonical Wegner generator, which requires a clear separation of energy scales in order to successfully diagonalise the full Hamiltonian. Fig.~\ref{fig.scramble2}b) shows the relative error in the (many-body) eigenvalues $\varepsilon$ for an interacting system of size $L=10$, with $\Delta_0=1.0$ (i.e., beyond the regime in which our truncation scheme can be expected to be accurate). Despite these strong interactions, the Wegner generator used in tandem with the scrambling transform returns surprisingly accurate results even at very weak disorder, while the Wegner generator alone performs extremely poorly at low disorder strengths due to its inability to remove off-diagonal terms which couple near-degenerate single particle sectors, leading in turn to the divergence of the interacting part of the Hamiltonian. At larger values of the disorder strength $d$, both methods return almost identical results.

\subsection*{Numerical considerations}

\begin{figure}
    \centering
    \includegraphics[width=0.45\linewidth]{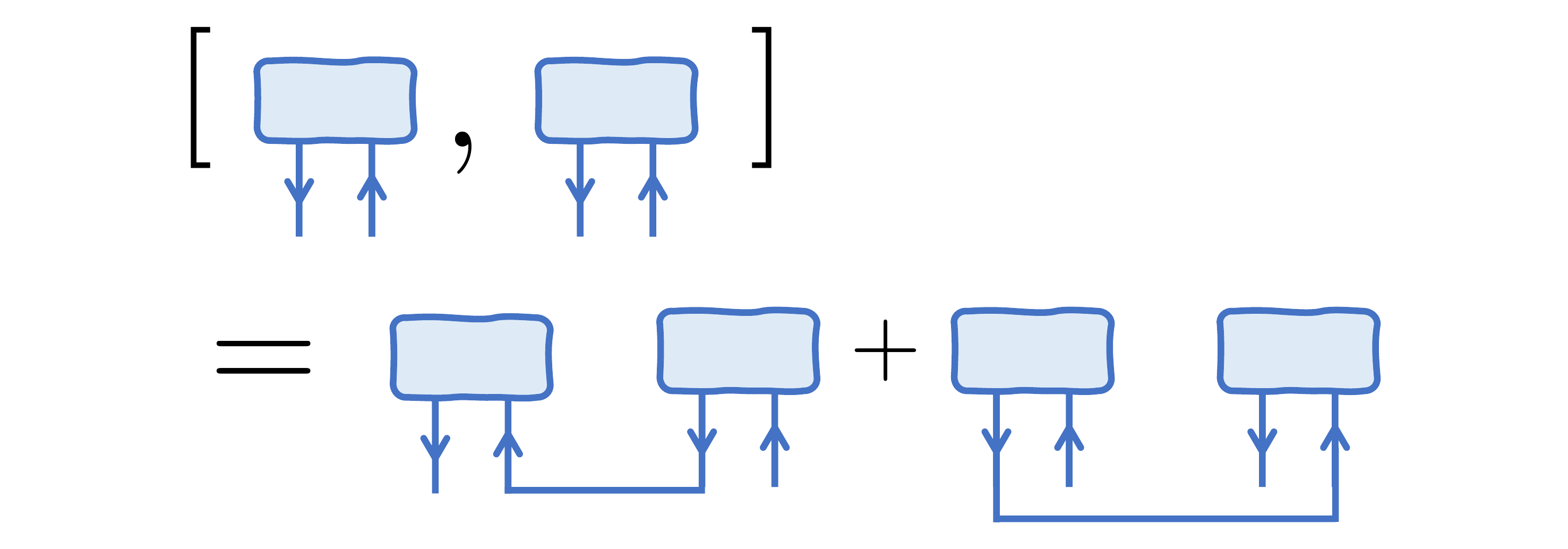}
    \includegraphics[width=0.45\linewidth]{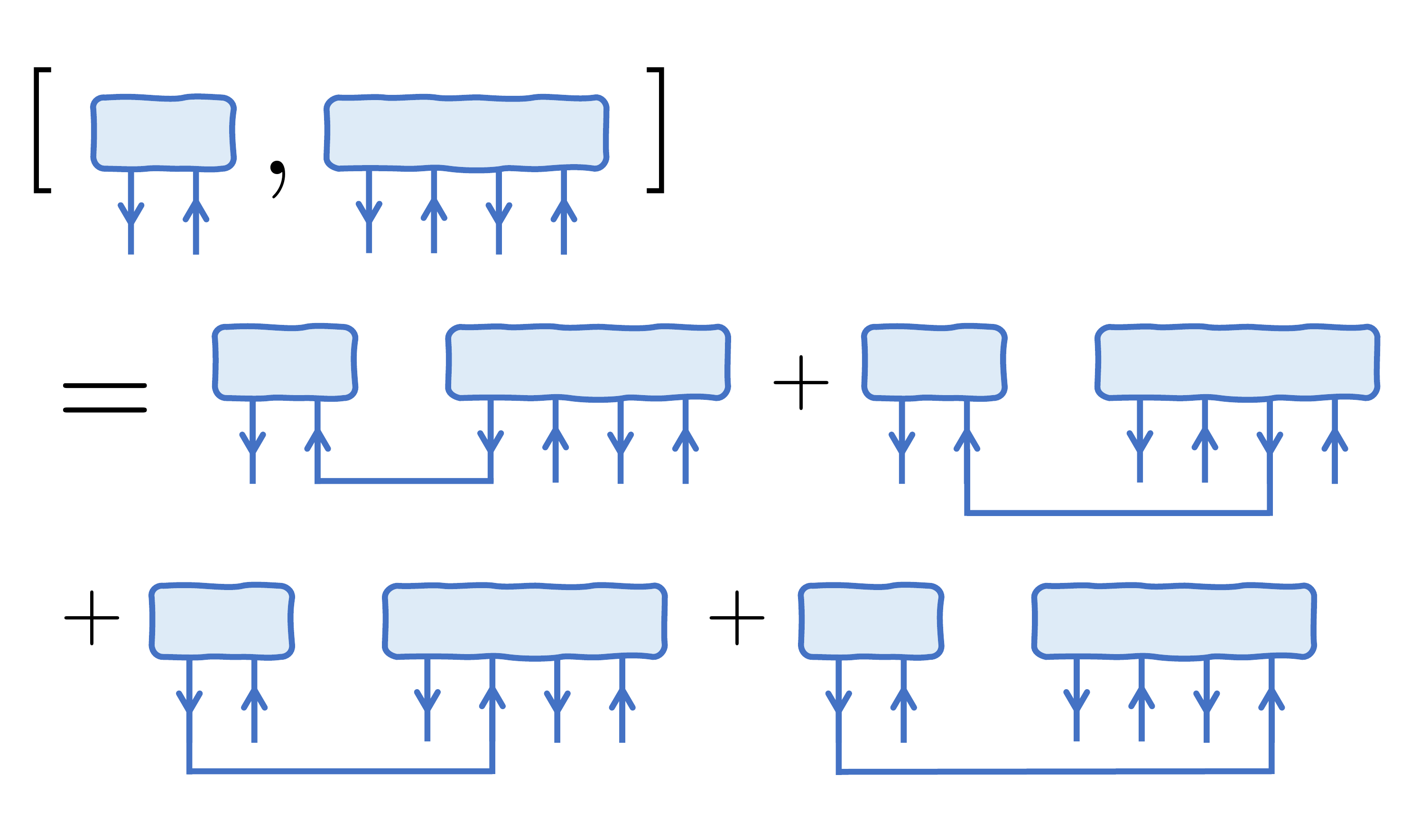}
    \caption{a) A schematic of how the commutator between two matrices may be computed in terms of one-point contractions. Creation and annihilation operators are indicated by up and down arrows respectively, and connected indices are summed over. The indices of the resulting arrays are rearranged into a consistent order (taking into account the appropriate (anti)commutation relations), and then added together to get the final result. This is equivalent to $\sum_{i,j,k,q}A_{ij} B_{kq} [c^{\dagger}_i c_j,c^{\dagger}_k c_q] =\sum_{i,j,k} A_{ik}B_{kq} c^{\dagger}_k c_q + A_{ij}B_{ki} c_j c^{\dagger}_k$. b) A schematic of how the commutator between
    \je{an order two and four}
    array may be computed in terms of one-point contractions. The most important thing is that the indices are put in a consistent order -- using appropriate (anti)commutation relations -- before the four resulting arrays are added together.}
    \label{fig.comm22}
\end{figure}

In order to make this process efficient, we make use of the 
\emph{tensor flow equation} (TFE) technique which recasts the commutators as a series of matrix/tensor contractions, avoiding the algebraic complexity of previous methods by turning the construction of the flow equation into an efficient numerical procedure. Any Hamiltonian may be written in a generic form as
\begin{align}
    \je{H}(l) &
    \je{:=}
    \je{H}^{(2)}(l) + \je{H}^{(4)}(l) +... \nn \\ 
    &= \sum_{i,j} H_{ij}(l) + \sum_{i,j,k,q} H_{ijkq}(l) + ...
\end{align}
where the superscripts indicate how many operators appear in each term, e.g., $\je{H}^{(2)}(l)$ contains all quadratic terms, $\je{H}^{(4)}(l)$ all quartic terms and so on. This can be stored in memory as a matrix ($\je{H}^{(2)}$) and a fourth-order tensor ($\je{H}^{(4)}$),  plus higher-order terms as required. A similar decomposition can be performed for the generator
\begin{align}
    \eta(l)
    \je{:=} \eta^{(2)}(l) + \eta^{(4)}(l) 
    +... \,.
\end{align}
This means that the flow equation can be computed as
\begin{align}
    [\eta(l),H(l)] &= [\eta^{(2)}(l),H^{(2)}(l)] + [\eta^{(2)}(l),H^{(4)}(l)] + [\eta^{(4)}(l),H^{(2)}(l)] + ... \,.
    \label{eq.flow}
\end{align}
Each of these commutators can be rewritten as the sum of all one-point contractions between the arrays $\eta^{(n)}(l)$ and $H^{(m)}(l)$, as shown in Fig.~\ref{fig.comm22}a) and Fig.~\ref{fig.comm22}b), with the result being an array of order $(n+m-2)$~\cite{Thomson+23}. Note that if any arrays are present with an order greater than $2$, the above commutator cannot be written as a closed expression and some form of truncation must be employed.
The source term in the flow (Eq.~(\ref{eq.flow})) which generates sixth order terms is given by
\begin{align}
    &[\eta^{(4)}(l),\je{H}^{(4)}(l)] = [[\je{H}_0^{(4)}(l),V^{(2)}(l)] + [\je{H}_0^{(2)}(l),V^{(4)}(l)],\je{H}^{(4)}(l)]
\end{align}
the elements of which we can estimate to be of order $\je{O}(J_0 \Delta_0^2)$ or smaller. A more detailed error analysis is given in the main text.
For weak interactions, even the induced sixth-order terms will be extremely small, and the eighth-order and higher terms will be essentially negligible. The exception to that is if any elements of $\je{H}^{(4)}(l)$ become of order one during the flow, at which point the higher order terms may be required to retain accuracy. The chances of this happening are heavily suppressed by the scrambling transform. We shall see later that higher-order terms in this framework enter with increasing powers of the interaction strength, and so by working at weak interactions we can ensure that the higher-order terms are essentially negligible. 

Two-point and higher contractions can also be taken into account~\cite{Thomson23}, however, we shall not consider these in the present manuscript as they are only non-zero if a state other than the vacuum is chosen for the normal-ordering procedure. This can be important for the incorporation of sophisticated non-perturbative corrections~\cite{Thomson23} or for finite-temperature contributions~\cite{Kehrein07}, but in the present case we will take the simpler approach of using vacuum normal-ordering.

The computational cost of this technique is set by the number of lattice sites, and by the highest-order of terms included in Eq.~(\ref{eq.Hrun}). In this work, we shall consider running Hamiltonians with up to fourth order ($\je{O}(L^4)$) and sixth order ($\je{O}(L^6)$) terms included. Crucially, the overall computational cost does \emph{not} depend on the geometry or dimensionality of the lattice sites: for example, a one-dimensional system of size $L=64$ can be diagonalized with essentially the same computational cost and accuracy as a two-dimensional system of size $L=8 \times 8$, or a three dimensional system of size $L = 4 \times 4 \times4$. 

\subsection*{Error analysis}

\begin{figure}[ht]
    \centering
    \includegraphics[width = 0.75\linewidth]{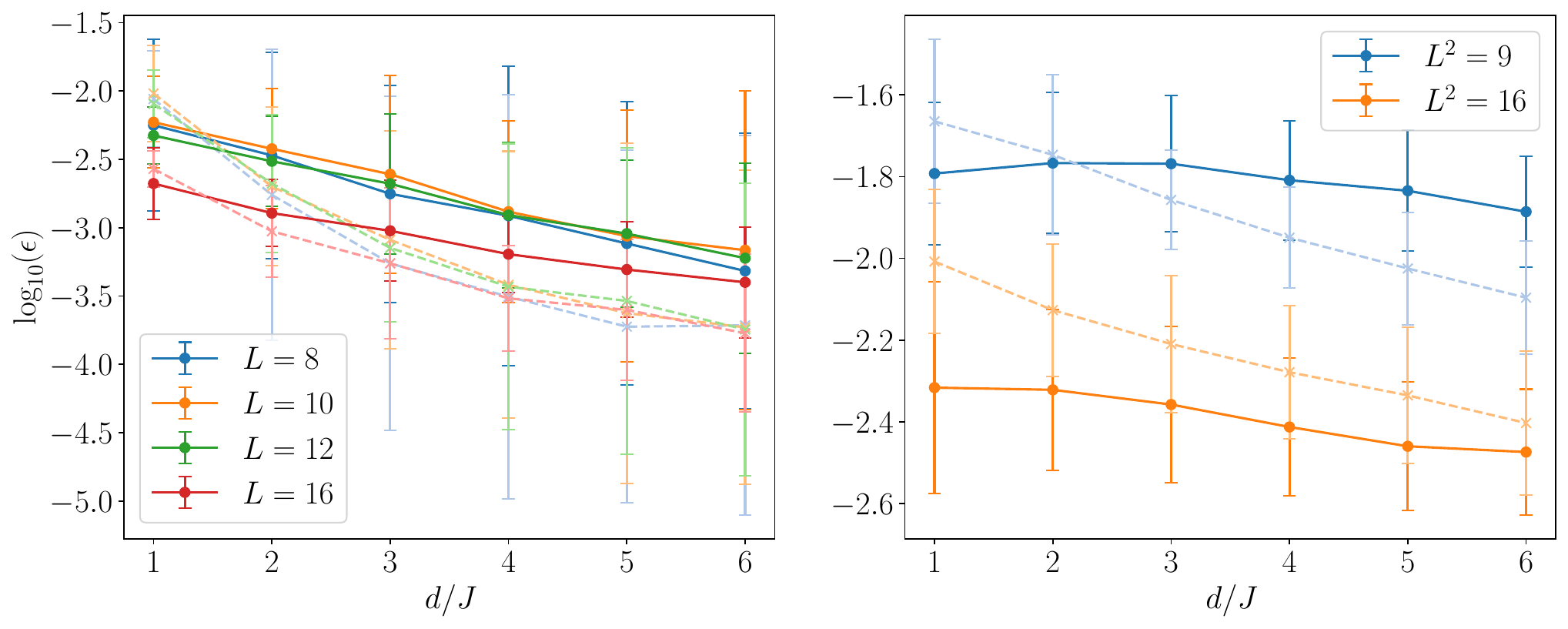}
    \caption{The relative error in the eigenvalues computed with flow equations and with exact digaonalisation, for a variety of system sizes and disorder strengths, averaged over $N_s \in [256,1024]$ disorder realizations depending on system size. a) The relative error in one dimension. b) The relative error in two dimensions.}
    \label{fig.ed_err}
\end{figure}

The relative error in the $i$-th eigenvalue $E_i$ is defined as
\begin{align}
    \epsilon_i := \left| \frac{E^{FE}_i - E^{ED}_i}{E^{ED}_i} \right|
\end{align}
where the superscripts $FE$ and $ED$ refer to the eigenvalues obtained by flow equations and exact diagonalisation respectively. We compute the median error within each disorder realization (as the error distribution has significant skewness, and a small number of outliers dominate the mean error), and then average this over disorder realizations to obtain the results shown in Fig.~\ref{fig.ed_err}. We find that the error remains small and under control for all system sizes and disorder strengths considered here. As expected, the error is larger in two dimensions than in one dimension, however, the error shrinks with increasing system size and remains within acceptable limits.

\begin{figure}[ht]
    \centering
    \includegraphics[width = 0.75\linewidth]{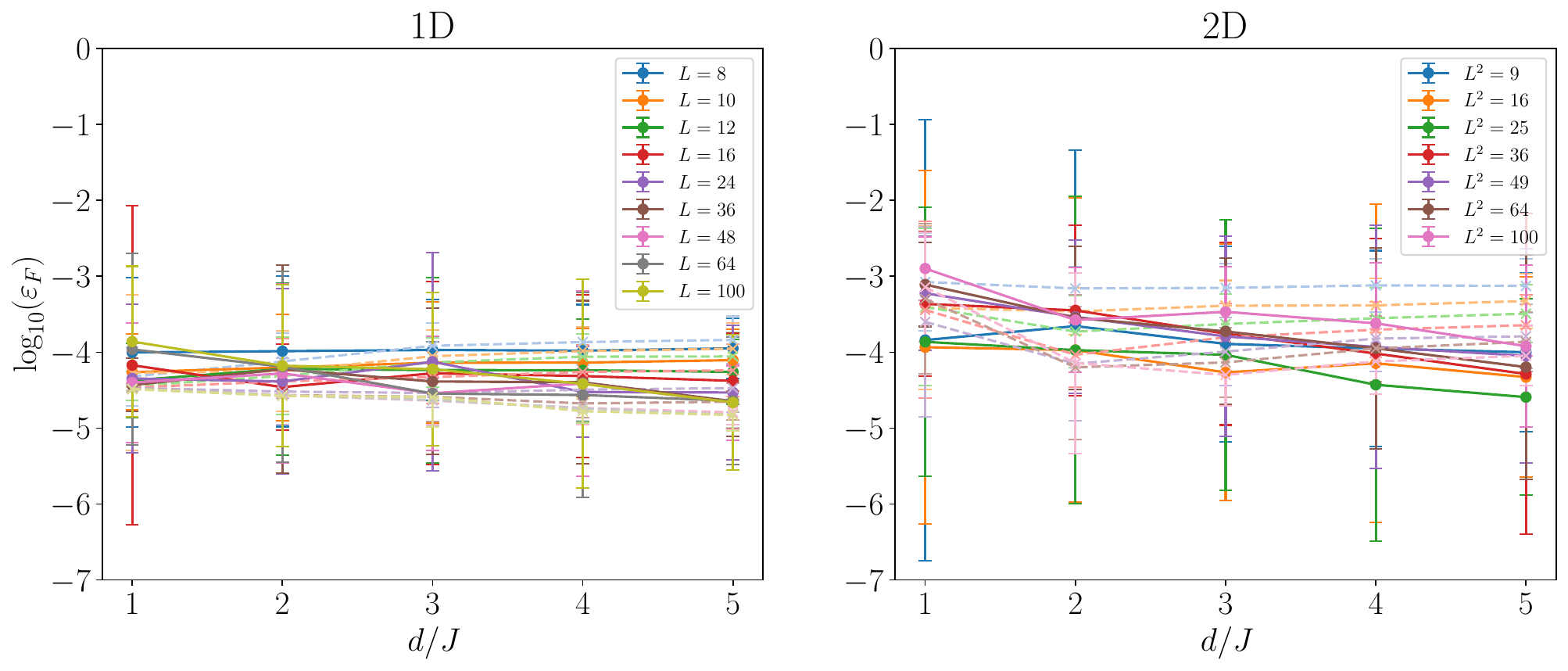}
    \caption{The average truncation error at each flow time step, as defined in the text, averaged over $N_s \in [32,1024]$ disorder realizations. Solid lines represent random disorder, and dashed lines show the results for the quasi-periodic potential. a) The truncation error in one dimension. b) The same quantity in two dimensions. Note that the error bars are much larger for random disorder than for the quasi-periodic potential, reflecting the greater likelihood of rare disorder-free regions (often known as `resonant regions' in the MBL literature) which act as effectively delocalised, and would require higher-order terms in the truncation to fully capture. These terms are typically absent in deterministic quasi-periodic potentials.}
    \label{fig.trunc_err}
\end{figure}

We can also define a second error metric, this time a self-consistent measure of the truncation error accrued over the course of the entire flow, based on a modified version 
\begin{align}
    \varepsilon_{F} = \frac{1}{l_{\textrm{max}}} \int_{0}^{l_{\textrm{max}}} \ud l  \|H^{(4)}_0(l)\|_F \|V^{(2)}(l)\|_F \|H^{(4)}(l)\|_F
\end{align}
of Eq.~(8) from the main text,
where the norms $\| . \|_F$ are now tensor Frobenius norms (i.e., square root of the sum of the squares of the entries of each tensor), and the integral is normalised by the total flow time, giving a measure of the average truncation error per timestep over the course of the entire diagonalization procedure. We approximate this integral numerically using the trapezoidal rule. Results are shown in Fig.~\ref{fig.trunc_err}, demonstrating that the errors due to the truncation of the running Hamiltonian remain small and under control throughout the procedure.

\je{Overall, there are two sources of errors: This is on the one hand 
the dominant truncation error
in the non-linear transformations along the fictitious time of the flow, 
bounded from above in Frobenius norm above. On the other hand, there is 
the error arising from pinching, when replacing matrices and tensors that are to a very
good approximation diagonal 
\begin{align}
    \je{H}(l_{\textrm{max}}
    ) = \sum_{i,j} H_{ij}(l_{\textrm{max}}) + \sum_{i,j,k,q} H_{ijkq}(l_{\textrm{max}}) 
\end{align}
for the final fictitious time $l_{\textrm{max}}$
by exactly diagonal matrices and tensors as
\begin{align}
    \tilde{\je{H}} &= \sum_{i\in {\cal L}} \tilde{h}_i :\tilde{n}_i: + \sum_{{i,j}\in {\cal L}} \Delta_{ij} :\tilde{n}_i \tilde{n}_j:  \,.
\end{align}
}

\subsection*{Computing the flow of operators}

Once we have diagonalized the Hamiltonian, we will wish to compute various observables in the diagonal basis, and so we need a way to transform arbitrary operators into the same basis as the Hamiltonian. This can be done straightforwardly, similarly to how the Hamiltonian itself is transformed. Rather than directly transforming fermionic number operators, as in previous works, here we will work directly with fermionic creation/annihilation operators. These are cheaper to compute by a factor of $L$, and once we know the form of either the creation or annihilation operator, the other follows straightforwardly by taking the Hermitian conjugate.

The flow of the creation operator is given by
\begin{align}
    \frac{\ud c^{\dagger}_i(l)}{\ud l} = [\eta(l),c^{\dagger}_i(l)].
\end{align}
which can be computed at the same time as the flow of the Hamiltonian, meaning that $\eta(l)$ does not need to be stored and can be discarded after each time step. Under the action of the flow, the creation operator takes the form
\begin{align}
    c^{\dagger}_i(l) = \sum_j A_j^{(i)} :\tilde{c}^{\dagger}_j: + \sum_{jkq} B_{jkq}^{(i)} :\tilde{c}^{\dagger}_j \tilde{c}^{\dagger}_k \tilde{c}_q: + ...
    \label{eq.tco}
\end{align}
which can be stored in memory as a unit order array (vector) plus 
an order three array (tensor), in contrast to the number operator which must be stored as a larger order two
array plus an order four array. To reconstruct the number operator at any point during the flow, we simply multiply Eq.~(\ref{eq.tco}) with its Hermitian conjugate using the definition 
\begin{equation}
n_i(l) := c^{\dagger}_i(l) c_i(l).
\end{equation}
Note that in addition to being cheaper by a factor of $L$ as compared with previous work which directly transformed the number operator up to quartic order~\cite{Thomson+23,Thomson23,Bertoni+22}, this also allows us immediate access to the sixth-order contributions to the number operator, which were not included in previous work. This approach is therefore faster, more efficient, and more accurate. In practice, we compute the creation operator to order $(O-1)$, where $O \in \{4,6\}$ is the order of the effective Hamiltonian. For $O=6$, this allows us to reconstruct the number operator up to terms including ten fermionic operators, in principle allowing us to take into account extremely weak effects that would only appear at high orders in perturbation theory. In practice, we do not go beyond sixth order, as otherwise the calculation becomes too memory-intensive for the present hardware, and for the largest system sizes considered in this work we do not go beyond fourth order.

This also allows us to transform (product) states from the microscopic basis to the diagonal basis, and vice-versa. We can write a generic product state vector as $\ket{\psi} = \Pi_i c^{\dagger}_i \ket{0}$ and transform each of the creation operators individually, multiplying the resulting expressions together at the end to obtain the transformed state. If the series of unitary transforms used to diagonalise the Hamiltonian has been stored, one can take any basis state (which are simply product states of zeros and ones in the diagonal basis) and reverse the transform in order to construct any desired eigenstate in the original microscopic basis. The resulting expression is likely to be quite complex, however, and it is unclear whether this procedure can be made of practical use.

\subsection*{Time evolution in the diagonal basis}

In the basis where the Hamiltonian is diagonal, operators can be time-evolved using the Heisenberg \je{equations} of motion
\begin{align}
    \frac{\ud O}{\ud t} = i [\tilde{\je{H}},O].
\end{align}
This can again be handled on a term-by-term basis. For an operator $O$ containing an odd number of fermionic operators, one obtains the  expression
\begin{align}
    [\tilde{\je{H}},O] &= [\tilde{\je{H}}^{(2)},O^{(1)}] + [\tilde{\je{H}}^{(2)},O^{(3)}] + [\tilde{\je{H}}^{(4)},O^{(1)}] + [\tilde{\je{H}}^{(4)},O^{(3)}] + \dots\,.
\end{align}
While for an operator $O$ containing an even number of fermionic operators, one obtains
\begin{align}
    [\tilde{\je{H}},O] &= [\tilde{\je{H}}^{(2)},O^{(2)}] + [\tilde{\je{H}}^{(4)},O^{(2)}] + [\tilde{\je{H}}^{(2)},O^{(4)}] + [\tilde{\je{H}}^{(4)},O^{(4)}] + \dots\,.
\end{align}
Once the operator has been time-evolved, we can compute its expectation value with respect to any basis state, or mixture of basis states. If the sequence of unitary transforms used to diagonalise the Hamiltonian has been stored, the transform can also be reversed in order to obtain the time-evolved operator back in the original basis. This is useful if one wishes to study operator spreading in real time, or compute the non-equilibrium dynamics starting from an initial state that is particularly simple in the original basis, e.g., some form of unentangled state like a Néel state.

\je{To be specific, we} will be particularly interested in computing time-evolved number operators. The number operators take the form
\begin{align}
    n_i(t,l) &= \sum_j \alpha^{(i)}_j(t,l) :\tilde{n}_j: + \sum_{j \neq k} \beta^{(i)}_{jk}(t,l) :\tilde{c}^{\dagger}_j \tilde{c}_k: \nn \\
    & \quad + \sum_{i,j,k,q} \Gamma_{ijkq}(t,l) :\tilde{c}^{\dagger}_i \tilde{c}_j \tilde{c}^{\dagger}_k \tilde{c}_q: + \sum_{i,j,k,q,l,m} \zeta_{ijkqlm}(t,l) :\tilde{c}^{\dagger}_i \tilde{c}_j \tilde{c}^{\dagger}_k \tilde{c}_q \tilde{c}^{\dagger}_l \tilde{c}_m: + \dots\, .
    \label{eq.TNO}
\end{align}
Higher order terms can be included, however, for the systems considered in this work they do not lead to any measurable difference in the results, and so we do not include them here.

In all of the following,  we drop the tilde notation for clarity, however, it is to be understood that from this point onwards, all operators are defined in the diagonal (tilde) basis. The number operator can now be time evolved according to the Heisenberg equation $\ud n_i(t)/ \ud t = i[\tilde{\je{H}},n_i(t)]$ and the contributions can be evaluated step-by-step. We could alternatively time-evolve the creation/annihilation operators, allowing us to avoid having to store the sixth-order term in computer memory, at the cost of having to perform a more complex series of summations when evaluating expectation values. The first contribution is given by the quadratic components of both the Hamiltonian and the number operator
\begin{align}
[\tilde{\je{H}}^{(2)},n_i^{(2)}] &= \sum_{ijk} \tilde{h}_k \beta_{i j} :[c^{\dagger}_k c_k:, :c^{\dagger}_i c_j:] \\
    &= \sum_{ijk}  \tilde{h}_k \beta_{i j} (\delta_{ik} - \delta_{jk}) :c^{\dagger}_i c_j: \\
    &= \sum_{ij} \beta_{i j} (h_i - h_j) :c^{\dagger}_i c_j:\,.
\end{align}
This is the only contribution to the quadratic part of the number operator, and it can be integrated to obtain
\begin{align}
    n^{(2)}_i(t) = \sum_j \alpha_j^{(i)} :n_j: + \sum_{j \neq k} \beta_{jk}^{(i)}(t) :c^{\dagger}_i c_j:
\end{align}
with \begin{align}
\beta_{jk}^{(i)}(t) := \exp(i (h_j - h_k)t) \beta_{jk}^{(i)}(0).
\end{align}

\je{The first term is constant, as the  terms that govern time evolution commute.}
There are two contributions to the quartic term. The first involves the (diagonal) quartic Hamiltonian
\begin{align}
    [\tilde{\je{H}}^{(4)},n_i^{(2)}] &= \sum_{i,j,k,q} \Delta_{ij} \beta_{kq} [:n_i n_j:,:c^{\dagger}_k c_q:] \nn \\
    &= \sum_{ijk} \Delta_{ij} :(\beta_{ik} c^{\dagger}_i c_k - \beta_{ki} c^{\dagger}_k c_i)c^{\dagger}_j c_j: + \sum_{ijk} \Delta_{ij} :c^{\dagger}_i c_i  (\beta_{jk} c^{\dagger}_j c_k - \beta_{kj} c^{\dagger}_k c_j):\,.
\end{align}
\je{The} second involves the quadratic part of the Hamiltonian and the quartic part of the number operator as
\begin{align}
    [\tilde{\je{H}}^{(2)},n_i^{(4)}] &= \sum_{ijkqm} \tilde{h}_m \Gamma_{ijkq} [:c^{\dagger}_m c_m:, :c^{\dagger}_i c_j c^{\dagger}_k c_q:] \nn \\
    &= \sum_{ijkq} (\tilde{h}_i-\tilde{h}_j+\tilde{h}_k-\tilde{h}_q) \Gamma_{ijkq} :c^{\dagger}_i c_j c^{\dagger}_k c_q:,
\end{align}
which together with the previous contribution implies 
\begin{align}
\Gamma_{ijkq}(t) = \exp[i(\tilde{h}_i-\tilde{h}_j+\tilde{h}_k-\tilde{h}_q)t]\Gamma_{ijkq}(0) + \delta_{ij} 
\frac{\Delta_{ik} (\beta_{kq}(t)-\beta_{kq}(0))} {\tilde{h}_k - \tilde{h}_q} + \delta_{kq} 
\frac{\Delta_{ik} (\beta_{ij}(t)-\beta_{ij}(0)}{\tilde{h}_i - \tilde{h}_j}.
\end{align}
\je{Once the Hamiltonian is in diagonal form,
one can also bound the error in the time evolution due to the imperfect diagonal
elements.
To upper bound the error in $\beta_{jk}^{(i)}(t) 
= \exp(i (h_j - h_k)t) \beta_{jk}^{(i)}(0)$: If the numbers $\{h_j\}$ are the approximations of the exact
$\{h_j'\}$ that are being used in the
time evolution in the above Hamiltonian, then the error made can easily be upper bounded as follows.
With 
\begin{align}
\delta:=
\sup_j |
h_j- h_j'| 
\end{align}
being the largest element-wise error, one obtains
\begin{eqnarray}
\left| \left(\exp(i (h_j' - h_k')t) - 
\exp(i (h_j - h_k)t) 
\right) \beta_{jk}^{(i)}(0) \right|
&=& \left|
\left(\exp(i (h_j' - h_k'- h_j+h_k)t
) - 
1
\right) 
\exp(i (h_j - h_k)t) 
\beta_{jk}^{(i)}(0)\right|\\
&\leq&
\frac{(h_j' - h_k'- h_j+h_k)^2 t^2}{2}|\exp(i (h_j - h_k)t)| 
\beta_{jk}^{(i)}(0)\nonumber\\
&\leq&
2\delta^2 t^2 
\beta_{jk}^{(i)}(0),
\nonumber
\end{eqnarray}
for all times $t\geq 0$. One can proceed similarly for the quartic term in the fermionic operators.}

In the very rare case of exact single-particle degeneracies, the $[\Delta_{ik} (\beta_{ij}(t)-\beta_{ij}(0))/(\tilde{h}_i - \tilde{h}_j)]$ and $[\Delta_{ik} (\beta_{kq}(t)-\beta_{kq}(0)) / (\tilde{h}_k - \tilde{h}_q)]$ terms are replaced by $\Delta_{ik}\beta_{ij}(0)t$ and $\Delta_{ik} \beta_{kq}(0)t$ respectively.
All higher-order terms can be evaluated similarly. As in the calculation of the commutators used in the flow equations, these can be efficiently computed using the graphical notation. In this work, we compute only the leading order phase shift applied to the sixth order terms coming from $[\tilde{\je{H}}^{(2)},n_i^{(6)}]$, which leads to 
\begin{equation}
\zeta_{ijkqlm}(t) = \exp[i(\tilde{h}_i-\tilde{h}_j+\tilde{h}_k-\tilde{h}_q + \tilde{h}_l - \tilde{h}_m)t] \zeta_{ijkqlm}(0). 
\end{equation}
Sixth-order terms are not included for system sizes $L>36$, as the memory cost becomes prohibitive, but the results for larger system sizes are entirely consistent with the trends observed for smaller systems, demonstrating that these terms have only a minor effect in all observed cases.
With this done, we have computed the time-evolved number operator at some arbitrary time \je{$t\geq0$}, without reference to any state, and without requiring step-by-step numerical evolution. Now we can move on and use this to compute observables.

\subsection*{Computing the infinite-temperature correlation function}

In previous works which have used flow equation techniques to compute the non-equilibrium dynamics of observables, the procedure has typically involved transforming the desired operator into the diagonal basis (where time evolution becomes easier), performing the time evolution, and then reversing the unitary transform to write the time-evolved operator back in the original basis, where appropriate expectation values could then be computed. This process was cumbersome, as it involved two unitary transforms (each with associated small but non-negligible numerical error), required storing the entire series of infinitesimal unitary transforms in memory, and the reverse transform had to be performed separately for each individual timestep. Altogether, this led to time evolution using flow equation techniques to be extremely computationally demanding, limiting their use.

Here, we demonstrate that infinite-temperature expectation values may be computed using a much more efficient approach.
The thermal expectation value of any arbitrary operator
$O$ \je{reflecting the Gibbs state} 
is given by 
\begin{equation}
\langle O \rangle = \textrm{Tr}[\exp(-\beta \je{H}) O]/\textrm{Tr}[\exp(-\beta \je{H})], 
\end{equation}
where $\beta =1/T$ is the inverse temperature (in units of $k_B=1$). In the infinite temperature limit, $\beta \to 0$, both exponentials reduce to the identity. In particular, the denominator simply becomes the trace of the identity, which gives the Hilbert space dimension, and the infinite-temperature expectation value becomes
\begin{align}
\langle O \rangle = \frac{1}{D_H} \sum_{i=0}^{D_H} \braket{\psi_i | O | \psi_i}
\label{eq.inf_temp}
\end{align}
where the $\{\ket{\psi_i}\}$ are the basis state vectors. This is simply the mean value of the expectation value of the operator $O$ taken over the entire Hilbert space.

While there are elegant ways to approximate this using dynamical typicality (where the sum over states is replaced by the expectation value in a single randomly chosen pure state), here we will take a simpler approach. We will replace the Hilbert space size $D_H$ with a smaller integer $N_s<D_H$, and approximate the mean value by
\begin{align}
\langle O \rangle \approx \frac{1}{N_s} \sum_{i=0}^{N_s} \braket{\psi_i | O | \psi_i}
\end{align}
For a sufficiently large value of $N_s$, this converges to the true value. In the following, we will choose $N_s = \textrm{min}(D_H,512)$ and restrict ourselves to the zero magnetization (half-filled) sector. We will work in the basis in which the Hamiltonian is diagonal, and the basis vectors are simply given by all possible binary strings $\ket{0,1,0,\dots}$. This allows us to avoid having to explicitly compute all eigenstates in the initial microscopic basis, which would be exponentially costly in the system size. This can be done by inserting the identity $U U^{\dagger} = \mathbbm{1}$ like so
\begin{align}
\langle O \rangle &\approx \frac{1}{N_s} \sum_{i=0}^{N_s} \braket{\psi_i |U U^{\dagger} O U U^{\dagger} | \psi_i}
\nonumber \\
&=\frac{1}{N_s} \sum_{i=0}^{N_s} \braket{\tilde{\psi}_i |\tilde{O} | \tilde{\psi}_i}
\label{eq.otrace}
\end{align}
reducing the problem to computing the transformed operator $\tilde{O} = U^{\dagger} O U$, and then computing its expectation value across $N_s$ product states. 

In this work, we are interested in computing the infinite-temperature autocorrelation function, defined as
\begin{align}
    C(t) = 4 \langle (n_i(t) - 1/2)(n_i(0) - 1/2) \rangle
\end{align}
where the factor of $4$ is chosen for convenience such that $C_i(0)=1$, and where $n_i(t)$ is obtained as previously described. We shall choose site $i$ to be in the middle of the system in all the following so that we are far from the boundaries and finite-size effects should be minimized. 
We will benchmark the performance of this approximation by comparison with exact diagonalization, making use of dynamical quantum typicality~\cite{Goldstein+06,Reimann07,Bartsch+09,Bartsch+11,Elsayed+13,Richter+19,Heitmann+20,Chiaracane+21} to efficiently compute the infinite temperature correlation function with an error exponentially small in the Hilbert space dimension $D_H$. In this case, the trace over basis states will be replaced by the expectation value taken with respect to a single state 
\je{vector}
\begin{align}
\ket{\psi} = C \sum_k (a_k + i b_k) \ket{\phi_k}, 
\end{align}
where \je{$C>0$} is a normalization constant, $a_k$ and $b_k$ are random real variables chosen from Gaussian distributions of zero mean and variance $1/2$. The state is chosen randomly for each disorder realization. This is essentially a way of approximating the trace in Eq.~\je{(\ref{eq.inf_temp})} by 
\je{uniformly}
randomly sampling contributions from all basis states,
\begin{align}
    & \langle (:n^{(2)}(t):+:n^{(4)}:(t))(:n^{(2)}(0):+:n^{(4)}(0):) \rangle \nn \\
    &= \langle :n^{(2)}(t): :n^{(2)}(0):\rangle + \langle :n^{(2)}(t): :n^{(4)}(0): \rangle \nn \\
    & \quad + \langle :n^{(4)}(t)::n^{(2)}(0):\rangle + \langle :n^{(4)}(t)::n^{(4)}(0): \rangle .
\end{align}
We can address each of these terms individually. 
In computing these expectation values, we must remember that the fermionic operators that we are using are normal-ordered with respect to the vacuum, purely to ensure a consistent ordering of operators is used throughout the flow procedure. To straightforwardly take the expectation value with respect to an arbitrary state, we can undo the $:\dots:$ notation and write the operators in explicitly normal-ordered form. We could of course have done this at any stage of the procedure, however, i) this formalism allows for more general normal-ordering procedures to be employed in future work, and ii) writing everything explicitly in vacuum normal-ordered form introduces additional minus signs that need to be carefully taken care of. We need the identities
\begin{align}
    :c^{\dagger}_i c_j: &= c^{\dagger}_i c_j ,\\
    :c^{\dagger}_i c_j c^{\dagger}_k c_q: &= c^{\dagger}_i c_j c^{\dagger}_k c_q - \delta_{jk} c^{\dagger}_i c_q = - c^{\dagger}_i 
    c^{\dagger}_k c_j c_q,
\end{align}
which are valid for vacuum normal-ordering and follow from the more general cases shown in Ref.~\cite{Kehrein07}. 
For the quadratic terms, this leads to
\begin{align}
    &\langle :n^{(2)}(t): :n^{(2)}(0):\rangle = \sum_{j,k} \alpha^{(i)}_j \alpha^{(i)}_k \langle n_j n_k \rangle + \sum_{j,k,l,m} \beta^{(i)}_{j,k}(t) \beta^{(i)}_{lm}(0) \langle c^{\dagger}_j c_k c^{\dagger}_l c_m \rangle \rangle.
\end{align}
Each of the expectation values can be computed by the application of Wick's theorem to determine which contractions are non-zero, keeping careful track of the minus signs that arise when swapping the order of operators. As we are taking expectation values with respect to a product state, this leads to
\begin{align}
    &\langle :n^{(2)}(t): :n^{(2)}(0):\rangle = \sum_{j,k} \alpha^{(i)}_j \alpha^{(i)}_j \langle n_j \rangle \langle n_k \rangle + \sum_{j,k} \beta^{(i)}_{jk}(t) \beta^{(i)}_{kj}(0) \langle n_j \rangle (1 - \langle n_k \rangle).
\end{align}
For the higher-order terms, we proceed as
\begin{align}
    &\langle :n^{(2)}(t): :n^{(4)}(0): \rangle =  -\sum_{j,p,q,l,m} \alpha^{(i)}_{j} \Gamma_{pqlm}(0) \langle c^{\dagger}_j c_j c^{\dagger}_p c^{\dagger}_l c_q c_m\rangle - \sum_{j,k,p,q,l,m} \beta^{(i)}_{jk}(t) \Gamma_{pqlm}(0) \langle c^{\dagger}_j c_k c^{\dagger}_p c^{\dagger}_l c_q c_m \rangle.
\end{align}
For convenience, we can rearrange the operators, to give
\begin{align}
    \langle c^{\dagger}_j c_k c^{\dagger}_p c^{\dagger}_l c_q c_m \rangle &= \delta_{kp} \langle c^{\dagger}_j  c^{\dagger}_l c_q c_m \rangle - \langle c^{\dagger}_j c^{\dagger}_p c_kc^{\dagger}_l c_q c_m \rangle \nn \\
    &= \delta_{kp} \langle c^{\dagger}_j  c^{\dagger}_l c_q c_m \rangle - \delta_{lk} \langle c^{\dagger}_j c^{\dagger}_p c_q c_m \rangle + \langle c^{\dagger}_j c^{\dagger}_p c^{\dagger}_l c_k c_q c_m \rangle .
\end{align}
This allows us to extract the lower-order contributions immediately, and now we only need to worry about the Wick contractions at each order.
For example, the contractions for the quartic term give
\begin{align}
\langle c^{\dagger}_j c^{\dagger}_l c_q c_m \rangle &= \langle \wick{\c1 c^{\dagger}_j \c2 c^{\dagger}_l \c2 c_q \c1 c_m}\rangle  + 
\langle \wick{\c1 c^{\dagger}_j \c2 c^{\dagger}_l \c1 c_q \c2 c_m}\rangle \nn \\
&= \langle c^{\dagger}_j c_m \rangle \langle c^{\dagger}_l c_q \rangle - \langle c^{\dagger}_j c_q \rangle \langle c^{\dagger}_l c_m \rangle \nn \\
&= \delta_{jm} \delta_{lq} \langle n_j \rangle \langle n_q \rangle - \delta_{j,q} \delta_{lm} \langle n_j \rangle \langle n_l \rangle.
\end{align}
Similarly, the other sixth-order term gives
\begin{align}
    &\langle :n^{(2)}(t): :n^{(4)}(0): \rangle = -\sum_{j,p,q,l,m} \alpha^{(i)}_{j} \Gamma_{pqlm}(t) \langle c^{\dagger}_p c^{\dagger}_l c_q c_m c^{\dagger}_j c_j \rangle - \sum_{j,k,p,q,l,m} \beta^{(i)}_{jk}(0) \Gamma_{pqlm}(t) \langle  c^{\dagger}_p c^{\dagger}_l c_q c_m c^{\dagger}_j c_k\rangle.
\end{align}
The operators can be rearranged in a similar way,
\begin{align}
    \langle  c^{\dagger}_p c^{\dagger}_l c_q c_m c^{\dagger}_j c_k\rangle &= \delta_{mj} \langle  c^{\dagger}_p c^{\dagger}_l c_q c_k\rangle - \langle  c^{\dagger}_p c^{\dagger}_l c_q c^{\dagger}_j c_m c_k\rangle \nn \\
    &= \delta_{mj} \langle  c^{\dagger}_p c^{\dagger}_l c_q c_k\rangle - \delta_{j,q} \langle  c^{\dagger}_p c^{\dagger}_l c_m c_k\rangle + \langle  c^{\dagger}_p c^{\dagger}_l c^{\dagger}_j c_q c_m c_k\rangle
\end{align}
and the Wick contractions computed as above. Finally, the eighth-order term can be written
as
\begin{align}
    &\langle :n^{(4)}(t): :n^{(4)}(0): \rangle =  \sum_{i_1,\dots ,i_8} \Gamma_{i_1 i_2 i_3 i_4}(t) \Gamma_{i_5 i_6 i_7 i_8} \langle c^{\dagger}_{i_1} c^{\dagger}_{i_3} c_{i_2} c_{i_4} c^{\dagger}_{i_5} c^{\dagger}_{i_7} c_{i_6} c_{i_8} \rangle,
\end{align}
which can likewise be rearranged into explicitly vacuum normal-ordered form and the expectation value computed by Wick's theorem. For small system sizes, the products of quadratic and sixth-order terms is also included in the numerics, but we do not go into further details here as they follow the same pattern as the previous terms.
Once we have all of these contributions, the expectation values in each state can be calculated and the sum in Eq.~(\ref{eq.otrace}) computed.

\end{document}